\documentclass[aoas,preprint]{imsart}

\RequirePackage[OT1]{fontenc}
\RequirePackage{amsthm,amsmath}
\RequirePackage{natbib}
\RequirePackage[colorlinks,citecolor=blue,urlcolor=blue]{hyperref}
\RequirePackage[hmarginratio=1:1,top=32mm,columnsep=20pt]{geometry} 
\RequirePackage{textgreek}
\RequirePackage{bm, bbm, blkarray, adjustbox}
\RequirePackage{stackengine}
\RequirePackage{amsfonts}
\RequirePackage{amssymb}
\RequirePackage{arydshln}
\RequirePackage{graphicx, float} 
\RequirePackage{scrextend}
\RequirePackage{algorithm,algorithmic}
\RequirePackage{ulem}
\RequirePackage{caption}
\RequirePackage{booktabs}

\begin{document}

\begin{frontmatter}
\title{Multivariate Bayesian structured variable selection for pharmacogenomic studies}
\runtitle{Structured variable selection}
\footnotetext{{\it $^*$Address for correspondence: }Department of Biostatistics, University of Oslo, P.O.Box 1122 Blindern 0317 Oslo, Norway. $^\dagger$These are joint last authors. \\ E-mail: zhi.zhao@medisin.uio.no}
\begin{aug}
\author{Zhi Zhao$^{* 1,2}$, Marco Banterle$^3$, Alex Lewin$^{\dagger 3}$, Manuela Zucknick$^{\dagger 1}$}

\affiliation{$^1$University of Oslo, Norway}
\affiliation{$^2$Oslo University Hospital, Norway}
\affiliation{$^3$London School of Hygiene \& Tropical Medicine, UK}

\end{aug}

\begin{abstract}
Precision cancer medicine aims to determine the optimal treatment for each patient. In-vitro cancer drug sensitivity screens combined with multi-omics characterization of the cancer cells have become an important tool to achieve this aim. Analyzing such pharmacogenomic studies requires flexible and efficient joint statistical models for associating drug sensitivity with high-dimensional multi-omics data. We propose a multivariate Bayesian structured variable selection model for sparse identification of omics features associated with multiple correlated drug responses. Since many anti-cancer drugs are designed for specific molecular targets, our approach makes use of known structure between responses and predictors, e.g. molecular pathways and related omics features targeted by specific drugs, via a Markov random field (MRF) prior for the latent indicator variables of the coefficients in sparse seemingly unrelated regression. The structure information included in the MRF prior can improve the model performance, i.e. variable selection and response prediction, compared to other common priors. In addition, we employ random effects to capture heterogeneity between cancer types in a pan-cancer setting. The proposed approach is validated by simulation studies and applied to the Genomics of Drug Sensitivity in Cancer data, which includes pharmacological profiling and multi-omics characterization of a large set of heterogeneous cell lines. 
\end{abstract}

\end{frontmatter}

\section{Introduction}\label{section:introduction}

A large proportion of advanced solid tumors harbor potentially treatable genomic variants \citep{Jardim2015,LeTourneau2015,VonHoff2010}, but very few cancer patients actually benefit from genome-informed treatments \citep{Marquart2018}. Thus, there is great potential to improve the use and benefit of therapy for individual patients by better patient stratification and by patient-tailored design of therapies. Precision cancer medicine aims at guiding cancer patient treatment based on detailed molecular characterization of each patient’s disease. One strategy that is rapidly gaining traction is ex vivo cancer drug sensitivity screening, which predicts responses to a range of potential therapies in cancer cell lines and patient-derived cells and identifies molecular features that are associated with drug response. Studies where both, drug sensitivity and molecular (multi-omics), data are available are commonly referred to as pharmacogenomic studies. In this article we employ a multivariate (multi-response) regression setup with high-dimensional input matrix to analyze pharmacogenomic data, where sensitivities to several drugs are the response variables and molecular (multi-)omics variables are the input features. We analyze data from the Genomics of Drug Sensitivity in Cancer (GDSC) database \citep{Garnett2012, Yang2013}, which contains the results from drug sensitivity screens to hundreds of cancer drugs for hundreds of cell lines representing diverse cancers in a pan-cancer setup and multi-omics characterization of these cell lines. Our approach can identify important genes affiliated with target pathways of the drugs (i.e. target genes) as well as genes whose dysfunction is known to drive cancer (cancer genes), which may guide personalized cancer therapies and aid discovery of potential new application areas of anti-cancer drugs in additional cancer types based on the identification of both tissue-specific and pan-cancer processes. 

Large-scale {\it in vitro} cancer drug screens produce a large amount of drug sensitivity data which are expected to be correlated for drugs that have similar mechanisms of action or common target genes or pathways. Meanwhile, multi-omics information, including for example transcriptomics (gene expression), genomics (point mutations or copy number variations) or epigenomics (e.g. CpG methylation) data, is measured for the cancer cells, which is expected to guide personalized cancer therapies through prediction of drug sensitivity \citep{Garnett2012, Barretina2012}. The omics input data are often high-dimensional and are typically sparsely associated with the response variables in a structured manner, where variables corresponding to genes in the same molecular pathway can have similar association patterns with the drugs, for example because a drug targets a molecular signaling pathway which effects the expression of several genes in the pathway. In addition, since multiple omics characterizations reflect different aspects' of information of the same system or co-functionality of multiple gene features \citep{Kim2019}, an analysis of joint associations between the correlated multiple phenotypes (e.g, multiple drugs) and high-dimensional molecular features (i.e. multi-omics data) is desired, but poses both theoretical and computational challenges. Finally, it is expected that not all of the heterogeneity between the cancer samples can be explained by the available molecular data. In particular, a pan-cancer pharmacogenomic screen will include samples from multiple cancer types, which adds heterogeneity in the drug sensitivity due to the different tissue and cell types, even if the involved molecular pathways and mechanisms are the same. This leads us to include random effects in the model to reflect heterogeneity between cancer types.

There are a number of statistical and machine learning models developed for predicting drug sensitivity by using omics data (see e.g. \cite{Ballester2022, Sharifi-Noghabi2021, Feng2021, Adam2020}). These models are often designed for making accurate predictions, either within a single cancer type \citep{Costello2014} or using a cancer-agnostic approach \citep{Barretina2012}. Furthermore, while emphasizing accurate predictions, many of the models lack effective variable selection options, making such black-box models less practical for biological studies or clinical applications. \cite{Huang2020} developed Tissue-guided LASSO for integrating cancer tissue of origin with genomic profiles, which just repeats the analysis in each cancer type, rather than jointly modelling the pan-cancer data. \cite{Zhao2020} proposed Tree-guided group lasso with Integrative Penalty Factors to jointly model drug-drug similarities and heterogeneity of multi-omics from pan-cancer data, but do not take into account correlation structure across multiple omics data sources.

Bayesian modeling provides flexibility to specify the relationships in such complex data. There have been several Bayesian methods developed to deal with structure in complex data. For example, \cite{Bai2020} and \cite{Yang2020} studied Bayesian group selection of high-dimensional predictors, but for univariate response variables. \cite{Liquet2017} extended the univariate response model to a multivariate model, but lack computational efficiency because they used a standard MCMC algorithm. \cite{Richardson2011} proposed hierarchical related regression (HRR) for multivariate response variables. HRR assumes a simple independence prior for the residual covariance matrix, and it applies an efficient Evolutionary Stochastic Search (ESS) algorithm based on Evolutionary Monte Carlo \citep{Bottolo2010}. More complex priors, e.g., inverse Wishart or hyper-inverse Wishart prior, can be used for the residual covariance matrix to learn structures between multivariate response variables \citep{Petretto2010, Carvalho2007, Wang2010, Bhadra2013, Bottolo2021}. 

Besides imposing different structured priors on the residual covariance matrix, it is necessary to also impose structured variable selection priors for high-dimensional predictors. 
Although independent spike-and-slab priors for variable selection are often used in high-dimensional multivariate models \citep{Jia2007, Bottolo2021, Ha2021, Chakraborty2021}, a structured MRF prior can also be used for the latent indicator variables to introduce prior dependence between predictors \citep{Chekouo2015, Chekouo2016, Chekouo2017}, and hyperpriors of the MRF prior can be used to infer the sparsity of the dependence structure. \cite{Lee2017} utilized the residual covariance matrix for the dependence structure in an MRF prior to encourage joint selection of the same predictor across several correlated response variables. In all these articles, an MRF prior is set for the latent variables of regression coefficients only corresponding to one response variable, which therefore does not allow to learn structures across multiple response variables.

\begin{figure}[H] 
\centering
\includegraphics[height=0.3 \textwidth]{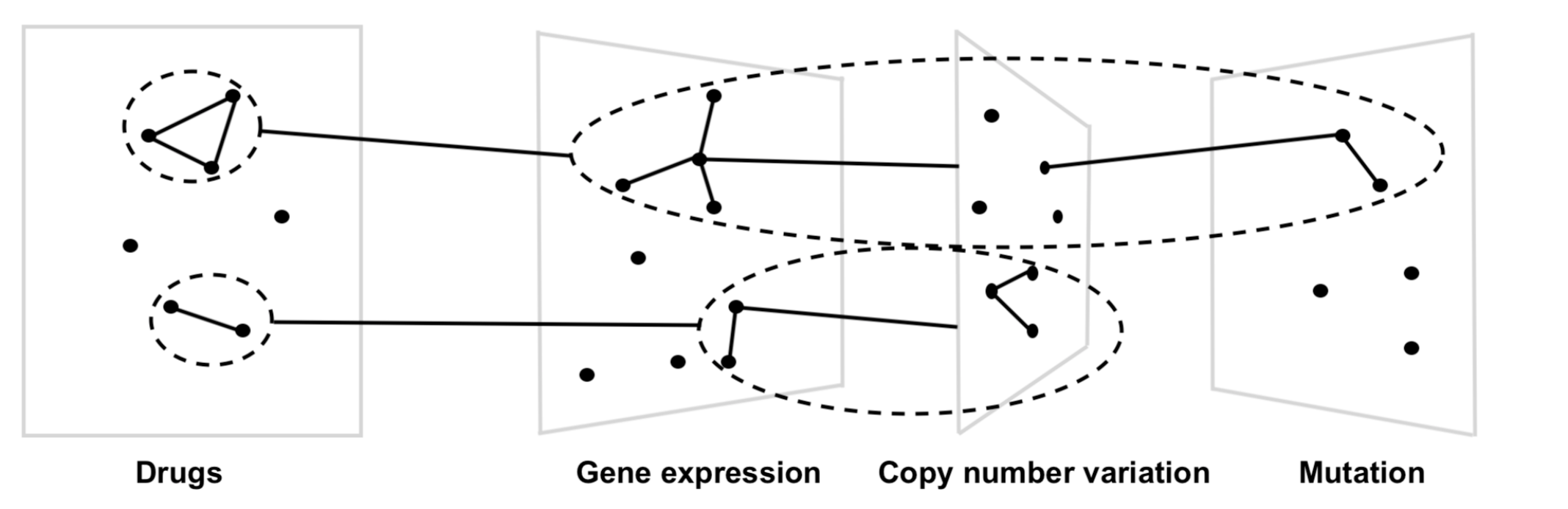}
\caption{Illustration of the drug groups and omics path (adapted from \cite{Ruffieux2019}).}
\label{fig:drugGene}
\end{figure}

In this article, we propose a multivariate Bayesian structured variable selection approach based on \cite{Richardson2011} and its extension by \cite{Bottolo2021}, which can deal with multiple response variables (e.g., the cell lines' sensitivity to multiple cancer drugs) and high-dimensional genomic predictors, and possess computational efficiency through the ESS algorithm. Our proposed approach aims to include a known complex structure between multiple response variables and high-dimensional predictors via a flexible MRF prior for the latent indicator variables of the regression coefficient matrix. That is, we include known biological associations for the dependence structure in an MRF prior rather than doing MRF inference. Our use of the MRF prior has two main advantages:
\begin{itemize}
\item it takes into account prior knowledge on inter-relations between predictors including across groups of predictors and across response variables, to improve model performance (i.e. variable selection and prediction), and
\item it performs posterior inference for the model in a more computationally efficient manner than the use of data-driven structured priors (e.g. multiplicative prior for the Bernoulli probability of the latent indicator variable (i.e. hotspot prior) by \cite{Richardson2011} and hyperprior for the MRF edge potentials by \cite{Chekouo2017}) would allow.
\end{itemize}
For example, Figure \ref{fig:drugGene} illustrates two groups of drugs and their corresponding two groups of target genes or pathways across multiple omics characterizations. When using omics data to predict drug responses, the associations between the multiple drugs and omics features can include prior knowledge about the groups of drugs and their target genes or target pathways. An MRF prior is able to address the joint structure by adding the edges for omics features within a group of target genes or pathways that correspond to the group of their targeting drugs. In addition, if the drug responses are measured on cell lines from different cancer types or different tissues, we use random effects to capture the sample heterogeneity arising from these sample groups. 
An R package BayesSUR \citep{Zhao2021} is available on the Comprehensive R Archive Network at \url{https://CRAN.R-project.org/package=BayesSUR.}

The rest of the article is organized as follows. In Section \ref{sec:models}, we introduce the Bayesian SSUR model,  propose an MRF prior for the latent indicator variables of the coefficient matrix, and introduce random effects for sample groups. Section \ref{sec:simulation} compares the performances of Bayesian SSUR models with our MRF prior to the hotspot prior by \cite{Bottolo2011} with respect to (w.r.t.) structure recovery and prediction in simulated data. In Section \ref{sec:GDSC}, we analyse a pharmacogenomic data set from the GDSC database. In Section \ref{sec:conclusion}, we conclude the article with a discussion.

\section{Methodology} \label{sec:models}

\subsection{SSUR model} \label{model:IW}
We study a multivariate regression model with a response matrix $\mathbf{Y}_{n \times m}$ from $n$ samples and $m$ response variables. All response variables are regressed on the same $p$ predictors which are measured on the $n$ samples, so that the predictor matrix is $\mathbf{X}_{n \times p}$. Associations between the responses $\mathbf{Y}$ and predictors $\mathbf{X}$ are captured by a coefficient matrix $\bm{B}_{p \times m}$. We first assume correlated response variables, but independent samples. Section \ref{model:RE} will then extend the model to allow for correlated samples. The classic SUR model is defined as
\begin{align*}
\begin{split}
\mathbf{Y} = \mathbf{X}\bm{B} + \mathbf{U}, \text{vec}\{\mathbf{U}\} \sim  \mathcal{N}(\mathbf{0},\ \Psi \otimes \mathbb{I}_n ),
\end{split} \tag{1}\label{formula:SUR}
\end{align*}
where the residuals have correlated columns with covariance $\Psi$ and independent rows, and $\text{vec}\{ \cdot \}$ is to vectorize a matrix by column. 

In the Bayesian framework, to efficiently sample from the posterior distribution of the regression coefficients from (\ref{formula:SUR}), 
\cite{Zellner2010} reparametrized the SUR model and proposed a direct Monte Carlo procedure. 
\cite{Bottolo2021} used the same reparametrization for the SUR model, but with an inverse Wishart prior $\Psi \sim  \mathcal{IW}(\nu, \tau \mathbb{I}_m)$. Briefly, then model (\ref{formula:SUR}) can be rewritten as
\begin{align*}
\begin{split}
\bm{y}_j = \mathbf{X}\bm{\beta}_j + \sum_{l<j}\bm{u}_l\rho_{jl} + \bm{\epsilon}_j, \bm{\epsilon}_j \sim  \mathcal{N}(\mathbf{0},\ \sigma_j^2\mathbb{I}_n),
\end{split} \tag{2}\label{formula:reparSUR}
\end{align*}
where $\bm{u}_l = \bm{y}_l - \mathbf{X}\bm{\beta}_l$. The reparametrized parameters $(\sigma_j^2,\rho_{jl})$ have priors 
\begin{align*}
\begin{split}
\sigma_j^2 \sim  \mathcal{IG}\left(\frac{\nu-m+2j-1}{2}, \frac{\tau}{2}\right), \rho_{jl} | \sigma_j^2 \sim  \mathcal{N}\left(0, \frac{\sigma_j^2}{\tau}\right), j>l,
\end{split} \tag{3}\label{formula:reparIW}
\end{align*}
where $v$ is fixed and $\tau \sim  \mathcal{G}amma(a_{\tau},b_{\tau})$. Note that the joint distribution $f(\mathbf{Y} | \mathbf{X} , \mathbf{B}, \Psi)$ is the same regardless of the order used for the decomposition since we are simply factorising it by chain-conditioning \citep{Bottolo2021}.

The reparametrization factorizes the likelihood across multiple response variables possible, which especially benefits high-dimensional response variables. If only a few of the $p$ predictor variables are assumed to be associated with any of the response variables, we use a latent indicator matrix $\bm\Gamma=\{\gamma_{kj}\}$ for variable selection. If $\gamma_{kj}=1$, then $\beta_{kj}\neq 0$ and the $k$th predictor is regarded as an associated predictor to the $j$th response variable; otherwise $\gamma_{kj}=0$ and $\beta_{kj}= 0$. Independent spike-and-slab priors \citep{George1993, Brown1998} for the regression coefficients 
can be used to find a small subset of predictors that explains the variability of $\mathbf{Y}$, for example:
\begin{equation*}
\beta_{kj}| \gamma_{kj},w \sim \gamma_{kj} \mathcal{N}(0,\ w) + (1-\gamma_{kj})\delta_0(\beta_{kj}),    \tag{4}\label{formula:spike_slab}\\
\end{equation*}
where $w \sim  \mathcal{IG}(a_w, b_w)$ and $\delta_0(\cdot)$ is the Dirac delta function. 

We may not only introduce sparsity to the high-dimensional coefficient matrix, but also sparsity to the precision matrix $\Psi^{-1}$, which implies that the residuals $\bm{u}_l = \bm{y}_l - \mathbf{X}\bm{\beta}_l$ and $\bm{u}_j = \bm{y}_j - \mathbf{X}\bm{\beta}_j$ for only a few pairs of response variables $l \neq j$ have non-zero partial correlations, assuming a multivariate normal distribution for the residuals. Such a sparse precision matrix can be conceptualized as a graph $ \mathcal{G}$, with nodes representing the residual variables $\bm{u}_l$, and edges between them corresponding to non-zero elements of the precision matrix. \cite{Bottolo2021} used a hyper-inverse Wishart prior for $\Psi$ instead of an inverse Wishart prior, i.e. 
\begin{equation*}
\Psi \sim  \mathcal{HIW}_{ \mathcal{G}}(\nu, \tau \mathbb{I}_m).
\tag{5}\label{formula:HIW}\\
\end{equation*}
It assumes an underlying decomposable graph $ \mathcal{G}$ between residuals. The HIW prior on decomposable graphs greatly enhances computational power since the parameters are updated within each clique and there is no computationally expensive normalisation constant to calculate. Since the fully Bayesian estimation procedure produces edges averaged over many different graphs, the posterior mean graph can well approximate non-decomposable graphs \citep{Fitch2014}. 
A sparse graph $ \mathcal{G}$ can result in sparse $\Psi^{-1}$. So \cite{Bottolo2021} specified a $ \mathcal{B}ernoulli(\eta)$ prior for each edge of the graph. Then a Binomial prior is on the cardinality edge-set
\begin{equation*} 
|\mathcal{E}| \sim  \mathcal{B}inomial( m(m-1)/2, \eta),
\tag{6}\label{formula:graphPrior}
\end{equation*}
where $\eta \sim  \mathcal{B}eta (a_{\eta}, b_{\eta})$ controls the sparsity of the graph. Based on (\ref{formula:HIW}) and (\ref{formula:graphPrior}), the parameters $\bm\sigma^2$ and $\bm\rho$ are indexed across the response variables of each clique of $ \mathcal{G}$ rather than all response variables. In addition to sparse covariance selection, \cite{Bottolo2021} also used sparse variable selection for the predictor variables via a hotspot prior (i.e. a multiplicative prior) for the hyper-parameter $\omega_{kj}$ in $\gamma_{kj} \sim  \mathcal{B}er(\omega_{kj})$. A guideline of prior specifications for the hyper-inverse Wishart prior and spike-and-slab prior can be found in Supplementary S1. 

\subsection{SSUR model with MRF prior} \label{model:MRF}
Figure \ref{fig:drugGene} illustrates known relationships between drug responses and genomic predictors. As an example, imagine a group of drugs with the same mechanism of action, where the response of a cancer cell to these drugs depends on a certain gene to be silenced. Gene silencing can either occur via a genomic alteration (deletion event), missense mutation, or another down-regulation of gene expression. It might thus be observable in one or several omics features, e.g. gene expression, copy number variation or mutation data. We may include such prior knowledge in the SUR model (\ref{formula:SUR}), instead of using independent or hotspot priors \citep{Richardson2011,Lewin2016,Bottolo2021}. 

We propose to use an MRF prior for the latent indicator vector $\bm{\gamma}= \text{vec}\{\bm{\Gamma}\}$ to address prior structure for the associations between response variables and predictors. The MRF prior is
\begin{equation*}
f(\bm{\gamma}|d,e,G) \propto \exp\{d\mathbbm{1}^\top\bm{\gamma}+ e\bm{\gamma}^\top G \bm{\gamma}\}, \tag{7}\label{formula:mrf}
\end{equation*}
where the scalar $d$ controls overall model sparsity, scalar $e$ determines the strength of the structure relationships between responses and predictors, and $G$ is an $mp \times mp$ (possibly weighted) adjacency matrix representing a graph to include prior structure knowledge. Term $d\mathbbm{1}^\top\bm{\gamma}$ in (\ref{formula:mrf}) can be generalized to $\bm d^\top \bm\gamma$, where the vector $\bm d$ will assign different relative contributions to the prior selection probabilities of the predictors. To specify the scalar $d$, we refer to \cite{Lee2017} by using log-odds of a rough model sparsity (i.e. proportion of nonzero regression coefficients). To specify $e$, \cite{Stingo2011} suggested a separate simulation from (\ref{formula:mrf}) over a grid of $e$ to detect the ``phase transition" value, and then specified a Beta prior on $e$. However, due to much computational cost in high-dimensional $\bm\gamma$, especially in multivariate regressions when searching some large values of $e$ resulting in very dense models, we first estimate the roughly largest $e$ and then use a grid search for $e$ to identify its optimal value for the model. See Supplementary S2 for more details.

For the $G$ matrix, we assign a positive edge potential $\{k+j(p-1),\ k'+j'(p-1)\}$-element if the latent indicator variables $\gamma_{kj}$ and $\gamma_{k'j'}$ are correlated. To illustrate the idea, we consider a simple case with three response variables (i.e. $\bm{y}_1$, $\bm{y}_2$ and $\bm{y}_3$) and four predictors (i.e. $\bm{x}_1$, $\bm{x}_2$, $\bm{x}_3$ and $\bm{x}_4$). When the predictors $\bm{x}_1$ and $\bm{x}_2$ are assumed \emph{a priori} to be associated with responses $\bm{y}_1$ and $\bm{y}_2$, and $\bm{x}_3$ and $\bm{x}_4$ are assumed to be associated with $\bm{y}_3$, then $G$ is a $12 \times 12$ matrix given by equation (\ref{formula:Gmat}). Any nonzero element in $G$ above can be any positive number which indicates a weight for the prior relationship between two latent indicator variables. Here for simplicity, we construct a symmetric $G$ matrix and assume all nonzero weights to be 1. 

\begin{equation}\tag{8}\label{formula:Gmat}
\scriptsize
G = 
\begin{blockarray}{ccccccccccccc}
 & \gamma_{11} & \gamma_{21} & \gamma_{31} & \gamma_{41} & \gamma_{12} & \gamma_{22} & \gamma_{32} & \gamma_{42} & \gamma_{13} & \gamma_{23} & \gamma_{33} & \gamma_{43}  \\
\begin{block}{c(cccccccccccc)}
  \gamma_{11} & 0  & 1  & 0  & 0  & 1  & 1  & 0  & 0  & 0  & 0  & 0  & 0   \\
 \gamma_{21}  & 1  & 0  & 0  & 0  & 1  & 1  & 0  & 0  & 0  & 0  & 0  & 0  \\
 \gamma_{31}  & 0  & 0  & 0  & 0  & 0  & 0  & 0  & 0  & 0  & 0  & 0  & 0  \\  
 \gamma_{41}  & 0  & 0  & 0  & 0  & 0  & 0  & 0  & 0  & 0  & 0  &  0 & 0  \\
 \gamma_{12}  & 1  & 1  & 0  & 0  & 0  & 1  & 0  & 0  & 0  & 0  & 0  & 0   \\
 \gamma_{22}  & 1  & 1  & 0  & 0  & 1  & 0  & 0  & 0  & 0  & 0  & 0  & 0   \\
 \gamma_{32}  & 0  & 0  & 0  & 0  & 0  & 0  & 0  & 0  & 0  & 0  & 0  & 0 \\ 
 \gamma_{42}  & 0  & 0  & 0  & 0  & 0  & 0  & 0  & 0  & 0  & 0  &  0 & 0  \\
 \gamma_{13}  & 0  & 0  & 0  & 0  & 0  & 0  & 0  & 0  & 0  & 0  & 0  & 0   \\
 \gamma_{23}  & 0  & 0  & 0  & 0  & 0  & 0  & 0  & 0  & 0  & 0  & 0  & 0  \\
 \gamma_{33}  & 0  & 0  & 0  & 0  & 0  & 0  & 0  & 0  & 0  & 0  &  0 & 1  \\
 \gamma_{43}  & 0  & 0  & 0  & 0  & 0  & 0  & 0  & 0  & 0  & 0  &  1 & 0  \\
\end{block}
\end{blockarray} \ .
\end{equation}

Note that we might not know all exact relationships between response variables and predictors, but we still formulate the matrix $G$ based on what we know. For example, if we only know relationships between response variables and relationships between predictors, we can aggregate these relationships by $G_y \otimes G_x - \mathbbm{I}$. Here we use $ - \mathbbm{I}$ to only allow zero diagonals in $G$, because nonzero diagonals are already captured by the term $d\mathbbm{1}^\top\bm{\gamma}$. For example, if we assume that $\bm{y}_1$ and $\bm{y}_2$ are related w.r.t. each predictor, $\bm{x}_1$ and $\bm{x}_2$ are related w.r.t. each response variable, and $\bm{x}_3$ and $\bm{x}_4$ are related w.r.t. each response variable, this translates into the following three Kronecker products. We can then aggregate them by aligning their coordinates into the full matrix $G$.

\[\scriptsize
\underbrace{G_y}_{\text{for } \bm{y}_1 \text{ and } \bm{y}_2} \otimes \underbrace{G_x}_{\text{for } \bm{x}_1, \bm{x}_2, \bm{x}_3 \text{ and } \bm{x}_4} - \ \mathbbm{I} = 
\begin{pmatrix}
1 & 1\\
1 & 1
\end{pmatrix}
\otimes 
\mathbbm{I}_{4}
- \mathbbm{I}_{8} .
\]

\[\scriptsize
\underbrace{G_y}_{\text{for } \bm{y}_1, \bm{y}_2 \text{ and } \bm{y}_3} \otimes \underbrace{G_x}_{\text{for } \bm{x}_1 \text{ and } \bm{x}_2} - \ \mathbbm{I} = 
\mathbbm{I}_{3}
\otimes 
\begin{pmatrix}
1 & 1\\
1 & 1
\end{pmatrix}
- \mathbbm{I}_{6} .
\]

\[\scriptsize
\underbrace{G_y}_{\text{for } \bm{y}_1, \bm{y}_2 \text{ and } \bm{y}_3} \otimes \underbrace{G_x}_{\text{for } \bm{x}_3 \text{ and } \bm{x}_4} - \ \mathbbm{I} = 
\mathbbm{I}_{3}
\otimes 
\begin{pmatrix}
1 & 1\\
1 & 1
\end{pmatrix}
- \mathbbm{I}_{6} .
\]

\subsection{SSUR model with MRF prior and random effects} \label{model:RE}
The SSUR model with hotspot prior in Section \ref{model:IW} and SSUR model with MRF prior in Section \ref{model:MRF} both assume independent and identically distributed samples conditional on the predictors. However, samples can be heterogeneous, especially in applications with large sample size. For example, large-scale drug screens may include cell line samples from different cancer tissue types. We address the heterogeneity of multiple sample groups by introducing random effects into the model similarly to \cite{Chekouo2015}. 


Let $\mathbf{Z}_{n \times T}$ be indicator variables representing $n$ samples from $T$ heterogeneous groups. We define an SUR model which includes spike-and-slab priors (\ref{formula:spike_slab}), hyper-inverse Wishart prior (\ref{formula:HIW}), MRF prior (\ref{formula:mrf}) and random effects, where the random effects $\bm{B}_0 = \{ \beta_{0,tj}: t=1,\cdots,T; j=1,\cdots,m\}$, and all priors above are mutually independent:
\begin{align*}
\begin{split}
\mathbf{Y} &= \mathbf{Z}\bm{B}_0 + \mathbf{X}\bm{B} + \mathbf{U}, \\
\beta_{0,tj}| w_0 &\sim  \mathcal{N}(0,\ w_0), \\
\beta_{kj}| \gamma_{kj},w &\sim \gamma_{kj} \mathcal{N}(0,\ w) + (1-\gamma_{kj})\delta_0(\beta_{kj}), \\
w_0 &\sim  \mathcal{IG}(a_{w_0}, b_{w_0}),\\
w &\sim  \mathcal{IG}(a_w, b_w),\\
\bm{\gamma}|d,e,G &\propto \exp\{d\mathbbm{1}^\top\bm{\gamma}+ e\bm{\gamma}^\top G \bm{\gamma}\},\\
\text{vec}\{\mathbf{U}\} &\sim  \mathcal{N}(\mathbf{0},\ \Psi \otimes \mathbb{I}_n ),\\
\Psi &\sim  \mathcal{HIW}_{ \mathcal{G}}(\nu, \tau \mathbb{I}_m),\\
\tau &\sim  \mathcal{G}amma(a_{\tau},b_{\tau}),\\
\end{split} \tag{9}\label{formula:SSUR_RE}
\end{align*}

Let us look into details of the random effects. For any $i$th sample and $j$th response variable, we have $y_{ij} = x_{i\cdot}^\top\bm{\beta}_j + z_{i\cdot}^\top\bm{\beta}_{0,j}  + u_{ij}$. For the $i$th sample, the covariance between the $j$th and $j'$th response variables is $\psi_{jj'}$ that is the $jj'$-element of $\Psi$, since
$$\mathbb{C}\text{ov}[y_{ij}, y_{ij'}] = \mathbb{C}\text{ov}[x_{i\cdot}^\top\bm{\beta}_j  + z_{i\cdot}^\top\bm{\beta}_{0,j} + u_{ij}, x_{i\cdot}^\top\bm{\beta}_{j'}  + z_{i\cdot}^\top\bm{\beta}_{0,j'} + u_{ij'}]= \mathbb{C}\text{ov}[u_{ij},u_{ij'}] = \psi_{jj'}.$$
Although the priors for the coefficients $\bm{B}_0$ and $\bm{B}$ in (\ref{formula:SSUR_RE}) do not provide any correlation between different responses for the same sample, the hyper-inverse Wishart prior on $\Psi$ models correlations between the response variables, and so does an inverse Wishart prior on $\Psi$. If we look at the reparametrization (3) from the inverse Wishart prior, or similarly from the hyper-inverse Wishart prior, correlations between the response variables are contained in the reparametrized parameter $\bm{\rho}$.

For the $j$th response variable, the covariance between the $i$th and $i'$th samples is
\begin{align*}
\mathbb{C}\text{ov}[y_{ij}, y_{i'j}] &= \mathbb{C}\text{ov}[x_{i\cdot}^\top\bm{\beta}_j  + z_{i\cdot}^\top\bm{\beta}_{0,j} + u_{ij}, x_{i'\cdot}^\top\bm{\beta}_j  + z_{i'\cdot}^\top\bm{\beta}_{0,j} + u_{i'j}] = w x_{i\cdot}^\top x_{i'\cdot} + w_0 z_{i\cdot}^\top z_{i'\cdot} \\
&= \begin{cases}
  w x_{i\cdot}^\top x_{i'\cdot}, & \text{if } i\text{th} \text{ and } i'\text{th samples belong to different groups } (z_{i\cdot} \neq z_{i'\cdot}), \\
  w x_{i\cdot}^\top x_{i'\cdot} + w_0, & \text{if } i\text{th} \text{ and } i'\text{th samples belong to the same group } (z_{i\cdot} = z_{i'\cdot}),
\end{cases}
\end{align*}
in which the hyper-parameter $w_0$ in the random effect determines the correlation between two samples from the same group.

We would like a weakly informative prior for $\beta_{0,tj}$ based on previous studies or expert knowledge in applications. In pharmacogenomic studies from multiple cancer tissues, for predict drug responses a tissue effect is usually stronger than a gene effect. Therefore it is appropriate to specify a larger hyper-parameter $w_0$ than $w$. 

 
\subsection{Posterior computation} \label{model:conditionals}
Posterior inference for the SSUR model with the MRF prior with or without additional random effects can be done in a similar manner to \cite{Bottolo2021}. For the SUR model (\ref{formula:reparSUR}) with a hyper-inverse Wishart prior for the residual covariance matrix $\Psi$ and an MRF prior for the latent indicator variables $\bm\gamma$, the joint posterior distribution is
\begin{align*}
\tag{10} \label{formula:jointPost}
&f(\bm{B},\bm{\Gamma}, w, \bm{\rho}, \bm{\sigma}^2, \tau,  \mathcal{G}, \eta |\mathbf{Y},\mathbf{X}) \\ 
= &f(\mathbf{Y}| \mathbf{X}, \bm{B}, \bm{\rho}, \bm{\sigma}^2) f(\bm{B}| \bm{\Gamma}, w) f(\bm{\Gamma}| G,d,e) f(w) f(\bm{\rho}| \bm{\sigma}^2, \tau,  \mathcal{G}) f(\bm{\sigma}^2| \tau,  \mathcal{G}) f(\tau) f( \mathcal{G}|\eta) f(\eta) \\
= &\prod_{j}f(\bm{y}_{j}| \mathbf{X}, \bm{B}, \bm{\rho}, \bm{\sigma}^2) \prod_{k,j}f(\beta_{kj} | \gamma_{kj},w) f(\bm{\gamma}| G,d,e) f(w) \prod_{j,l<j} f(\rho_{jl}|\sigma_j^2, \tau, \mathcal{G}) \prod_{j}f(\sigma_j^2| \tau, \mathcal{G}) f(\tau) f( \mathcal{G}|\eta) f(\eta) ,
\end{align*}
where $\bm{\rho}$ and $\bm{\sigma}^2$ are vectors of $\{\rho_{jl}\}$ and $\{\sigma_j^2\}$, respectively. Since $\bm{y}_{j}| \mathbf{X}, \bm{B}, \bm{\rho}, \bm{\sigma}^2$ is normally distributed with mean $\mathbf{X}\bm{\beta}_j + \sum_{l<j}\bm{u}_l\rho_{jl}$ and variance $\sigma_j^2\mathbb{I}_n$, we can obtain the full conditional distributions of the regression coefficients $\bm{\beta}_j$, $w$, $\sigma_j^2$, $\rho_{jl}$ and $\tau$. The posterior distribution of the latent indicator variable $\bm\gamma=\text{vec}\{\bm{\Gamma}\}$ is estimated by a Metropolis-Hastings sampler. The graph $ \mathcal{G}$ of the hyper-inverse Wishart prior is sampled from a junction tree sampler which is essentially Metropolis-Hastings sampling \citep{Green2013}, see \cite{Bottolo2021} for more details.
If there are random effects for sample groups as in (\ref{formula:SSUR_RE}), the joint posterior distribution (\ref{formula:jointPost}) includes parameters $\bm{B}_0$ and $w_0$, i.e.
\begin{align*}
&f(\bm{B}_0,\bm{B},\bm{\Gamma}, w_0,w, \bm{\rho}, \bm{\sigma}^2, \tau,  \mathcal{G}, \eta |\mathbf{Y},\mathbf{X},\mathbf{Z})\\
= &f(\mathbf{Y}| \mathbf{X},\mathbf{Z}, \bm{B}_0, \bm{B}, \bm{\rho}, \bm{\sigma}^2) f(\bm{B}_0|w_0)f(w_0)  
f(\bm{B}| \bm{\Gamma}, w) f(\bm{\Gamma}| G,d,e) f(w)f(\bm{\rho}| \bm{\sigma}^2, \tau,  \mathcal{G}) f(\bm{\sigma}^2| \tau,  \mathcal{G}) f(\tau) f( \mathcal{G}|\eta) f(\eta). 
\end{align*}

We implement Gibbs samplers to obtain posterior estimates for $\bm{B}$, $\bm{\rho}$ and $\bm{\sigma}^2$, and update the latent indicator variable $\bm{\Gamma}$ via a Metropolis-Hastings sampler with parallel tempering in the same way as \cite{Bottolo2021}. Thompson sampling \citep{Russo2018} is used to derive the proposal for each latent indicator $\gamma_{kj}$. The hyper-parameter $\tau$ is updated via a random walk Metropolis sampler as proposed by \cite{Bottolo2021}. To overcome the prohibitive computational time in high-dimensional settings, the ESS algorithm \citep{Bottolo2010, Richardson2011} is used to update the posteriors. For each iteration of the MCMC sampler, after sampling the latent indicator variables $\bm{\Gamma}$, we first update the hyper-parameters ($\tau, w, w_0, \mathcal{G}$), then update the parameters $\bm{\sigma}^2$ and $\bm{\rho}$, and finally the regression coefficient matrices $\bm{B}$ and $\bm{B}_0$ (see Supplementary S3). At each iteration, the ESS algorithm implements a local move to add/delete and swap the latent indicator variables within each chain, and then a global move to exchange and crossover the latent indicators between any two parallel tempered chains. The temperature across all response variables is adapted based on the acceptance rate of the global exchange operator. The ESS algorithm with parallel tempering is effective in searching a high-dimensional model space with multiple modes \citep{Bottolo2010}.

\subsection{Model performance evaluation}\label{sec:evaluation}
To evaluate the performance of the proposed approach, we focus on structure recovery and prediction performance. The structure recovery includes the estimation of the latent indicator variable $\bm{\Gamma}$ which captures the relationships between response variables and high-dimensional predictors, and the estimation of the graph $ \mathcal{G}$ which addresses the residual relationships between response variables. Predictive accuracy of Bayesian models for new data points can be measured by the expected log pointwise predictive density (elpd), which can be assessed by leave-one-out cross-validation ($\text{elpd}_{\text{loo}}$) or by the widely applicable information criterion ($\text{elpd}_{\text{waic}}$) \citep{Vehtari2017}. We also calculate the  root mean squared prediction error (RMSPE) measured on an independent test data set for the median probability model (MPM) \citep{Barbieri2004, Barbieri2021} in addition to the training data root mean squared error (RMSE). 

\cite{Vehtari2017} proposed an efficient computation for the Bayesian LOO estimate of out-of-sample predictive fit
$\text{elpd}_{\text{loo}} = \sum_{j=1}^m\sum_{i=1}^n \log f(y_{ij} |\bm{y}_{(-i)j}),$
where $\bm{y}_{(-i)j}$ is the observation vector of the $j$th response variable except the $i$th observation. The LOO is estimated by
$$\widehat{\text{elpd}}_{\text{loo}} = \sum_{j=1}^m\sum_{i=1}^n \frac{1}{\frac{1}{N}\sum_{t=1}^N \frac{1}{f(y_{ij} |\bm{\vartheta}^{(t)})} },$$
where $N$ is the length of an MCMC chain and $\bm{\vartheta}^{(t)}$ is the MCMC samples at the $t$th iteration for all related parameters. The WAIC is estimated by
$$\widehat{\text{elpd}}_{\text{waic}} = \widehat{\text{elpd}}_{\text{loo}} - \sum_{j=1}^m\sum_{i=1}^n \mathbb{V}\text{ar}_{t=1}^N[\log f(y_{ij} |\bm{\vartheta}^{(t)})],$$
where the second term above is used as a measure of the model complexity.

\noindent For future prediction, a single model may be required in some cases, for practical reasons or for simplicity. 
\cite{Barbieri2004} suggested the median probability model (MPM), which is defined for each coefficient to be $\mathbbm{E}[\beta_{kj} | \gamma_{kj}=1, data]$ if $\mathbbm{P}\{\gamma_{kj}=1| data\}>0.5$, or 0 otherwise. It can be estimated through MCMC estimates:
$$\hat{\beta}_{kj, MPM}= 
\begin{cases}
  \frac{\sum_{t=1}^N \beta_{kj}^{(t)}}{\sum_{t=1}^N \gamma_{kj}^{(t)}}, & \text{if } \frac{\sum_{t=1}^N \gamma_{kj}^{(t)}}{N} > 0.5, \\
  0, & \text{otherwise},
\end{cases}$$
where $\gamma_{kj}^{(t)}$ is the estimate of the $t$th MCMC iteration for the latent indicator variable of $\beta_{kj}$. After obtaining $\hat{\bm{B}}_{MPM}=\{\hat{\beta}_{kj, MPM}\}$, the 
$$\text{RMSE} = \frac{1}{\sqrt{ mn}}\|\mathbf{Y} - \mathbf{X} \hat{\bm{B}}_{MPM}\|_2,$$
$$\text{RMSPE} = \frac{1}{\sqrt{ mn'}}\|\mathbf{Y}^* - \mathbf{X}^* \hat{\bm{B}}_{MPM}\|_2,$$
where $\mathbf{Y}_{n \times m}$ and $\mathbf{X}_{n \times p}$ were used to estimate $\hat{\bm{B}}_{MPM}$, and $\mathbf{Y}_{n' \times m}^*$ and $\mathbf{X}_{n' \times p}^*$ are new data.

\section{Simulation study}\label{sec:simulation}

\subsection*{}

In this section, our SSUR model with MRF prior, denoted as SSUR-MRF, is evaluated w.r.t. structure recovery for the regression coefficient matrix and prediction performance of responses. We set up two simulation scenarios: one with independent samples and the other with heterogeneous and correlated samples. In the first scenario, our approach is compared with the SSUR model with a hotspot prior, denoted as SSUR-hotspot, which was studied by \cite{Bottolo2021}. In the second scenario, our approach is compared with the SSUR-MRF model without random effects.

 
\subsection{Simulation scenarios}
\begin{figure}[H]
\centering
\includegraphics[height=0.35 \textwidth]{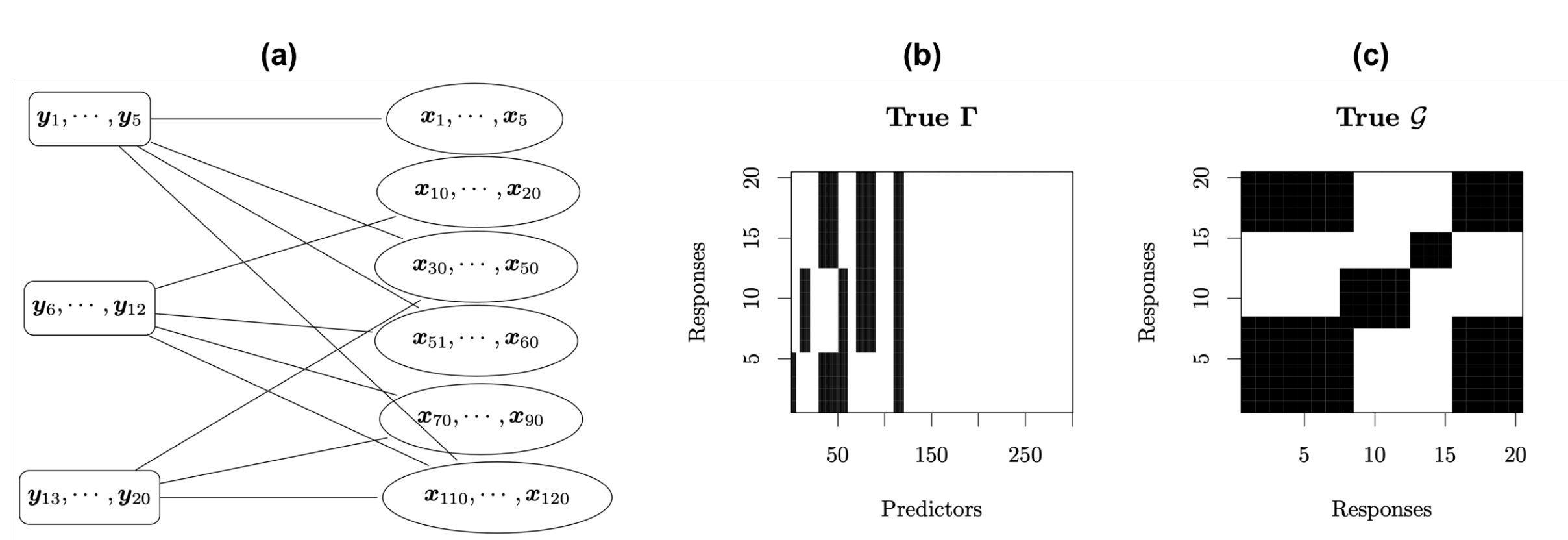}
\caption{Simulation scenarios: True relationships between response variables and predictors. (a) Network structure between $\mathbf{Y}$ and $\mathbf{X}$; (b) latent indicator variable $\bm{\Gamma}$ for the associations between $\mathbf{Y}$ and $\mathbf{X}$ in the SUR model; (c) additional structure $\mathcal{G}$ between response variables not explained by $\mathbf{X}\bm{B}$. Black indicates a true relation between the response variables and predictors.}
\label{fig:paramTrue}
\end{figure}

We design a network (Figure \ref{fig:paramTrue}(a)) to construct a complex structure between 20 response variables and 300 predictors. It assumes that the responses are divided into three groups, and the first 120 predictors are divided into six groups. The first group of responses ($\{\bm{y}_1,\cdots,\bm{y}_5\}$) is related to four groups of predictors ($\{\bm{x}_{1},\cdots, \bm{x}_{5}\}$, $\{\bm{x}_{30},\cdots, \bm{x}_{50}\}$, $\{\bm{x}_{51},\cdots, \bm{x}_{60}\}$ and $\{\bm{x}_{110},\cdots, \bm{x}_{120}\}$). The second group of responses ($\{\bm{y}_6,\cdots,\bm{y}_{12}\}$) is also related to four predictor groups ($\{\bm{x}_{10},\cdots, \bm{x}_{20}\}$, $\{\bm{x}_{51},\cdots, \bm{x}_{60}\}$, $\{\bm{x}_{70},\cdots, \bm{x}_{90}\}$ and $\{\bm{x}_{110},\cdots, \bm{x}_{120}\}$). The third group of responses ($\{\bm{y}_{13},\cdots,\bm{y}_{20}\}$) is related to three predictor groups ($\{\bm{x}_{30},\cdots, \bm{x}_{50}\}$, $\{\bm{x}_{70},\cdots, \bm{x}_{90}\}$ and $\{\bm{x}_{110},\cdots, \bm{x}_{120}\}$). Corresponding to this network structure between responses and predictors, a sparse latent indicator variable $\bm{\Gamma}$ (Figure \ref{fig:paramTrue}(b)) reflects the associations between response variables and predictors in the SUR model (\ref{formula:SUR}). In addition, we design a decomposable graph $ \mathcal{G}$ (Figure \ref{fig:paramTrue}(c)) to reflect the residual structure between the response variables. The graph $ \mathcal{G}$ has six blocks representing six subgroups of responses that cannot be explained by the linear predictor $\mathbf{X}\bm{B}$, which makes the modeling more challenging. The information in $ \mathcal{G}$ is included in the residuals and can be expected to be recovered by statistical models.

In scenario 1, the response and predictor datasets are generated based on a multivariate linear regression model
$\mathbf{Y} = \mathbbm{1}\bm{\alpha}^\top+\mathbf{X}\bm{B}_{\bm{\Gamma}} + \mathbf{U}.$
The intercepts $\bm{\alpha}=\{\alpha_j\}$ and input data $\mathbf{X}=\{x_{ik}\}$ ($i=1,\cdots,250; k=1,\cdots,300; j=1,\cdots,20$) are simulated independently from the standard normal distribution. The regression coefficients $\bm{B}=\{\beta_{kj}\}$ ($k=1,\cdots,300; j=1,\cdots,20$) are also simulated independently from the standard normal distribution but truncated by the latent indicator variable $\bm{\Gamma}=\{\gamma_{kj}\}$, i.e. $\bm{B}_{\bm{\Gamma}}=\{\beta_{kj}\mathbbm{1}_{\gamma_{kj}=1}\}$. The noise matrix $\mathbf{U}$ is simulated based on the multivariate normal distributed $\tilde{\mathbf{U}}$ 
and a G-Wishart distribution 
\citep{Mohammadi2019}. We first simulate the G-Wishart distribution $P \sim  \mathcal{W}_{ \mathcal{G}}(3,M)$ where diagonals of $M$ are 1 and the off-diagonals are 0.5, and then use Cholesky decomposition $\text{chol}(P^{-1})$ to obtain the noise matrix $\mathbf{U} = \tilde{\mathbf{U}} \cdot \text{chol}(P^{-1})$. We control the average signal-to-noise ratio defined by \cite{Bottolo2021}, and set it to 10. Independent data sets $\mathbf{X}^*$ and $\mathbf{Y}^*$ are simulated based on the same scenario as validation data. 
In scenario 2, $\mathbf{X}$, $\bm{\Gamma}$, $\bm{B}_{\bm{\Gamma}}$ and $\mathbf{U}$ are generated in the same manner as scenario 1. We also include group indicators $\mathbf{Z}$ with independent row vectors $\bm{z}_{i} \sim multinomial(0.1,0.2,0.3,0.4)$ ($i=1,\cdots,n$ and the number of groups is set to $T=4$), and random effects $\bm{B}_0$ 
with each group effect from $\mathcal{N}(0,2^2)$. 
The response dataset is generated from a linear mixed model
$\mathbf{Y} = \mathbf{X}\bm{B}_{\bm{\Gamma}}+ \mathbf{Z}\bm{B}_0 + \mathbf{U}.$
Independent validation data sets are also simulated based on scenario 2. The algorithms for the two simulation scenarios can be found in Supplementary S4. 
%
Both simulation algorithms generate validation datasets independently with the same sample size to evaluate the performance of the proposed methods. 

\subsection{Comparison of the SSUR-hotspot and SSUR-MRF models}
\begin{figure}[H]
\centering
\makebox[\textwidth][c]{\includegraphics[height=0.3 \textwidth]{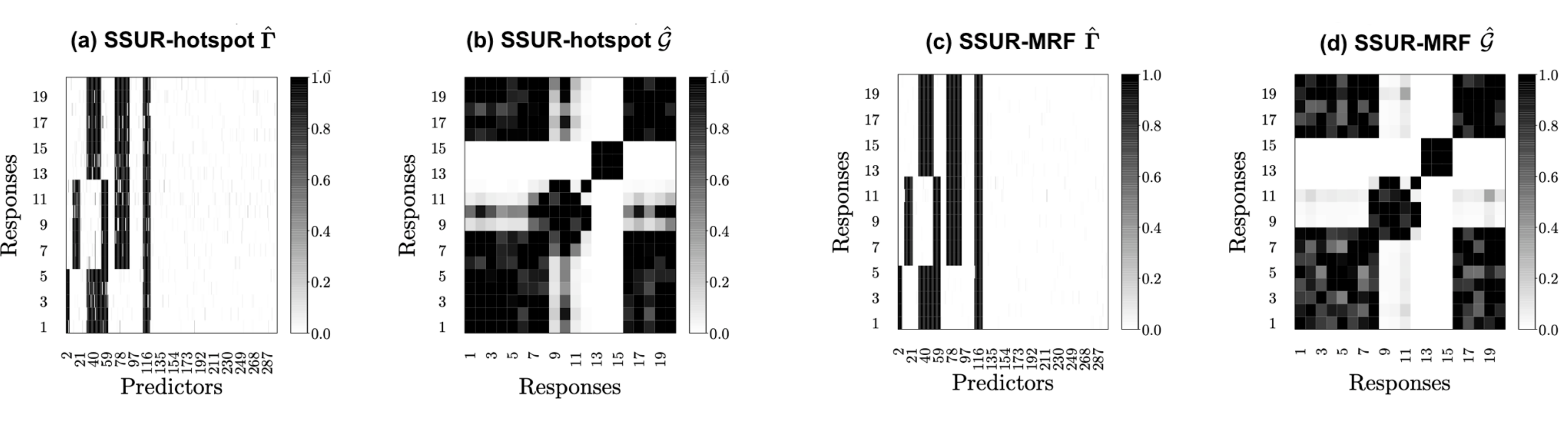}}
\caption{Results for simulation scenario 1: Posterior mean of $\bm\Gamma$ and $\mathcal{G}$ by models SSUR-hotspot (panels (a) and (b)) and SSUR-MRF (panels (c) and (d)) }
\label{fig:paramEst}
\end{figure}

We first compare our proposed SSUR-MRF model to the SSUR-hotspot model on simulated data generated with scenario 1.
Our approach uses the network in Figure \ref{fig:paramTrue}(a) as prior information to construct edge potentials for the MRF prior as illustrated in Section \ref{model:MRF}. Throughout this article, we refer to a predictor as being selected or identified, if the corresponding latent indicator variable has posterior mean larger than 0.5. Figure \ref{fig:paramEst} shows that SSUR-hotspot and our SSUR-MRF both have good recovery for the residual structure between response variables (i.e. $ \mathcal{G}$). However, SSUR-MRF has better structure recovery of the latent indicator variable $\bm{\Gamma}$. Table \ref{tab:paramEst} reports higher accuracy, sensitivity and specificity of the estimator for $\bm{\Gamma}$ by SSUR-MRF than SSUR-hotspot. The two methods have similar $\widehat{\text{elpd}}_{\text{loo}}$ and $\widehat{\text{elpd}}_{\text{waic}}$, but our approach has smaller RMSE and RMSPE.

\begin{table}[H]
\centering
\caption{Results for simulation scenario 1: Accuracy of variable selection and prediction performance of models SSUR-hotspot and SSUR-MRF prior} 
\label{tab:paramEst}
{
\begin{tabular}{ccccccccc} 
\toprule
\smallskip
 & accuracy & sensitivity & specificity & 
 RMSE & RMSPE \\ \hline
\smallskip
SSUR-hotspot & 0.988 & 0.936 & 0.999 & 0.800 & 0.693    \\ 
SSUR-MRF     & 0.989 & 0.998 & 0.986  & 0.643 & 0.412  \\ 
\toprule
\end{tabular}
}
\end{table}

\subsection{Sensitivity analysis for SSUR-MRF}
The MRF prior can be strongly informative as the edge potentials were constructed according to the true relationships $\bm{\Gamma}$ between the simulated response variables and predictors. Here we use different edge potentials $G$ in the MRF prior for a sensitivity analysis. Starting from the previously constructed edge potentials $G$, we partially delete true edge potentials, either uniformly or non-uniformly, or add noise edges, or aggregate Kronecker products between the three response groups and six predictor groups as shown in Figure \ref{fig:paramTrue}(a). The four cases are as follows:
\begin{itemize}
	\item \textbf{Case 1:} delete 1\%, 10\%, 50\% or 90\% edges uniformly from the fully informative $G$. This case for every block in $\bm{\Gamma}$, some corresponding edge potentials in $G$ are kept.
	\item \textbf{Case 2:} delete 1\%, 10\%, 50\%, 90\% or 100\% edges non-uniformly in consecutive chunks from the edge list\footnotemark
\footnotetext{The coordinates of all nonzero entries of $ \mathcal{G}$ are put in an edge list in order. Deleting edge potentials uniformly, e.g. deleting 1\%, means that the $1+(1-1/|\mathcal{E}|)/1\% \cdot \{0:(1\% \cdot |\mathcal{E}|)\}$th edges of the edge list are deleted. Deleting 1\% edges non-uniformly (i.e. blocks of edges) means that the last 1\% edges in the edge list are deleted. The edge list includes the edges of each pattern (i.e. association block) together. Adding 1\% noise edges means that $1\% \cdot mp(mp-1)/2$ wrong edges are included randomly.} of the fully informative $G$. In this case, for some blocks in $\bm{\Gamma}$ all corresponding edge potentials in $G$ are deleted.
	\item \textbf{Case 3:} add 0.1\%, 0.5\%, 1\%\footnotemark
	\footnotetext{Note that 1\% already exceeds the total number of true edges that are $\sim$ 0.3\% of all possible edges.} noise edges to the fully informative $G$.
	\item \textbf{Case 4:} aggregate Kronecker products between response groups and predictor groups (see guidance in Section \ref{model:MRF}).
\end{itemize}

\begin{figure}[H]
\centering
\makebox[\textwidth][c]{\includegraphics[height=0.28 \textwidth]{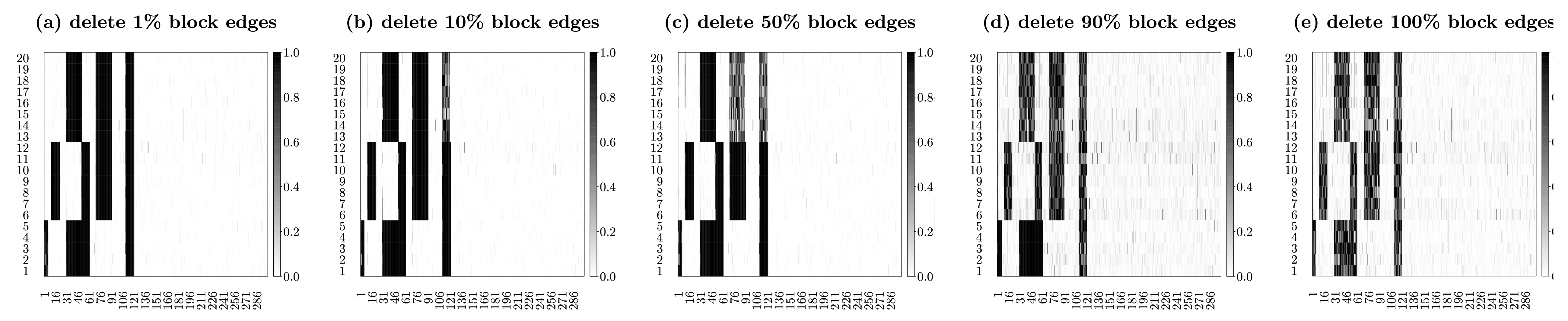}}
\caption{Results for simulation scenario 1: Sensitivity analysis for case 2, i.e. when blocks of edges are deleted (i.e. delete edges non-uniformly).}
\label{fig:sensitivityII}
\end{figure}

Table \ref{tab:sensitivity} Case 1 shows that our SSUR-MRF model can identify well truly associated predictors w.r.t. accuracy, sensitivity and specificity of the estimated $\bm\Gamma$, and have stable prediction performance w.r.t. RMSE and RMSPE, when deleting 1\%, 10\%, 50\% or 90\% true edges uniformly. This indicates that our approach can recover a good structure of $\bm\Gamma$ and good prediction performance of responses, even if only a little true association knowledge across all patterns of $\bm\Gamma$ is used in the MRF prior. Case 2 (Table \ref{tab:sensitivity}) where some of the patterns/blocks in $\bm\Gamma$ are fully unknown (i.e. when the corresponding blocks of edges in $G$ are deleted) in the MRF prior, the sensitivity of variable selection and prediction performance w.r.t. RMSE and RMSPE becomes slightly worse. Figure \ref{fig:sensitivityII} indicates that the information of the deleted blocks cannot be recovered fully, but will instead be estimated with a sparser $\bm{\Gamma}$. Supplementary S5 shows slightly worse residual structure recovery (i.e. $\hat{ \mathcal{G}}$) when deleting more edges non-uniformly. However, even the worst-case scenario in Case 2, when all edges are deleted, i.e. when the MRF prior with $G=\mathbf 0$ degenerates to a Bernoulli prior without any known structure information between variables, has similar performance to the SSUR model with hotspot prior in Table \ref{tab:paramEst}. Case 3 (Table \ref{tab:sensitivity}), where adding noise edges, shows similar variable selection and prediction performance to using true potential edges. Finally, Case 4 (Table \ref{tab:sensitivity}), where aggregating Kronecker products for the edge potentials in the MRF prior, the variable selection remains similar to using true potential edges. Here, $\widehat{\text{elpd}}_{\text{loo}}$ and $\widehat{\text{elpd}}_{\text{waic}}$ do not change much between different cases, but they can be used as the objective function to optimize hyperparameters.

\begin{table}[H]
\caption{Results for simulation scenario 1: Sensitivity analysis of SSUR-MRF with different MRF priors \label{Table2} } 
\label{tab:sensitivity}
\begin{adjustbox}{width=\textwidth}
\hfuzz=0.64pt
\begin{tabular}{rccccccc}
\toprule
\smallskip
                                                        & accuracy & sensitivity & specificity & $\widehat{\text{elpd}}_{\text{loo}}$ & $\widehat{\text{elpd}}_{\text{waic}}$ & RMSE & RMSPE \\ \cline{1-8} 
\multicolumn{1}{c}{\bf Case 1}  &   \\                                                                                                     
delete edges uniformly  &   \\ 
1\%      & 0.989 & 0.998 & 0.987 & -18612.2 & -18612.4 & 0.642 & 0.411 \\ 
10\%    & 0.989 & 0.998 & 0.987 & -18623.0 & -18624.1 & 0.642 & 0.411 \\ 
50\%    & 0.989 & 0.998 & 0.987 & -18618.7 & -18622.3 & 0.642 & 0.410\\ 
90\%    & 0.994 & 0.993 & 0.994 & -18622.1 & -18623.8 & 0.652 & 0.453  \\ 
\\
\multicolumn{1}{c}{\bf Case 2}  &   \\   
delete edges non-uniformly  &   \\ 
1\%                     & 0.989 & 0.998 & 0.987 & -18622.0 & -18623.8 & 0.642 & 0.412  \\ 
10\%                   & 0.988 & 0.989 & 0.988 & -18619.8 & -18620.6 & 0.668 & 0.470 \\ 
50\%                   & 0.986 & 0.959 & 0.991 & -18622.5 & -18624.2 & 0.806 & 0.723 \\ 
90\%  & 0.991 & 0.955 & 0.998 & -18622.6 & -18624.8 & 0.780 & 0.662 \\ 
100\% & 0.990 & 0.942 & 1.000 & -18623.1 & -18624.5 & 0.800 & 0.732 \\ 
\\
\multicolumn{1}{c}{\bf Case 3}  &   \\   
add noise edges  &   \\ 
0.1\%  & 0.989 & 0.998 & 0.987 & -18621.5 & -18623.8 & 0.643 & 0.414 \\  
0.5\%  & 0.989 & 0.998 & 0.987 & -18622.0 & -18623.4 & 0.643 & 0.413 \\ 
1\%     & 0.989 & 0.998 & 0.987 & -18621.2 & -18623.2 & 0.643 & 0.412 \\
\\ 
\multicolumn{1}{c}{\bf Case 4}  &   \\   
aggregate Kronecker products  & 0.990 & 0.998 & 0.988 & -18620.2 & -18623.0 & 0.644 & 0.412 \\ 
\toprule
\end{tabular}
\end{adjustbox}
\end{table}

\subsection{Results and discussion of SSUR-MRF with random effects}
\begin{figure}[H]
\centering
\makebox[\textwidth][c]{\includegraphics[height=0.3 \textwidth]{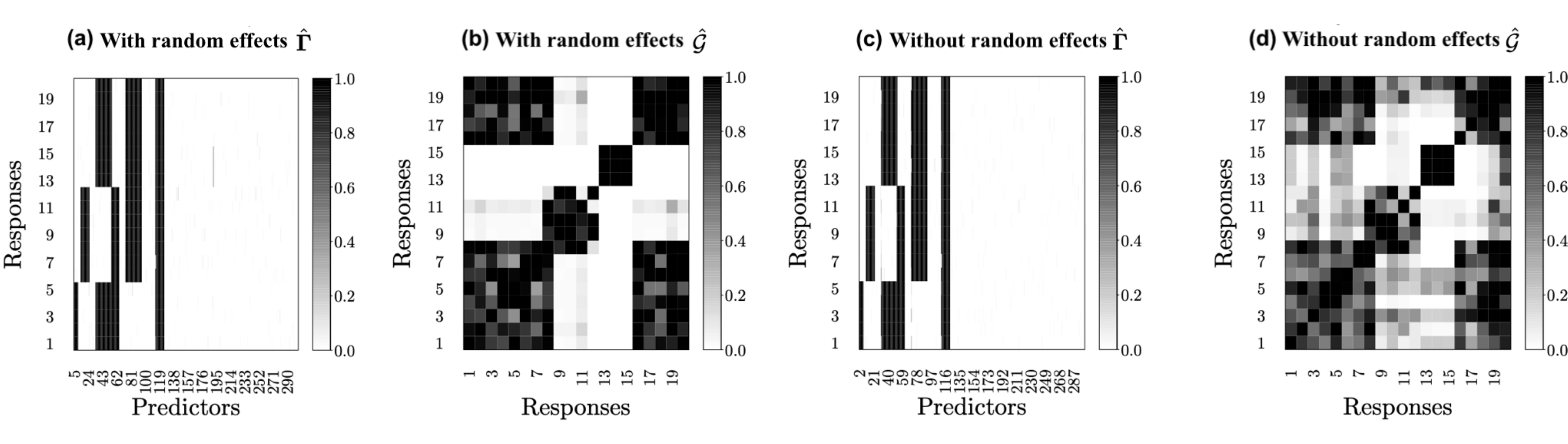}}
\caption{Results for simulation scenario 2: Posterior mean of $\bm{\Gamma}$ and $\mathcal{G}$ by the SSUR-MRF with random effects based on the simulated data from scenario 2}
\label{fig:paramEstRe}
\end{figure}

In the simulation scenario 2, $T=4$ sample group variables are simulated to assess the performance of our SSUR-MRF model with random effects. Figure \ref{fig:paramEstRe}(a), \ref{fig:paramEstRe}(c) and Table \ref{tab:paramEstRe} show similar recovery of the latent indicator variable $\bm{\Gamma}$ w.r.t. accuracy, sensitivity and specificity for both SSUR-MRF with and without random effects. However, SSUR-MRF model without random effects is difficult to recover the residual graph structure $ \mathcal{G}$ (Figure \ref{fig:paramEstRe}(d)), while the model with random effects can recover well the true structure (Figure \ref{fig:paramEstRe}(b)). See also Table \ref{tab:paramEstRe}, which reports the recovery performance of $ \mathcal{G}$ w.r.t. accuracy, sensitivity and specificity when thresholding its posterior mean at 0.5. For the response prediction, SSUR-MRF with random effects has smaller $\widehat{\text{elpd}}_{\text{loo}}$, $\widehat{\text{elpd}}_{\text{waic}}$, RMSE and RMSPE than SSUR-MRF without random effects (Table \ref{tab:paramEstRe}). In addition, for the accuracy of estimated regression coefficients, $\frac{1}{\sqrt{mp}}\|\hat{\bm{B}}_{MPM}-\bm{B}\|_{\ell_2}^2$ by SSUR-MRF without random effects has larger error (0.050) than by SSUR-MRF with random effects (0.013). In addition, the 

\begin{table}[H]
\centering
\caption{Results for simulation scenario 2: Structure recovery and prediction by SSUR-MRF with and without random effects \label{Table3} } 
\label{tab:paramEstRe}
\begin{adjustbox}{width=\textwidth}
\hfuzz=0.64pt
{\begin{tabular}{rccccccc}
\toprule
\smallskip
& accuracy &  sensitivity & specificity  & $\widehat{\text{elpd}}_{\text{loo}}$ & $\widehat{\text{elpd}}_{\text{waic}}$ & RMSE & RMSPE \\ \cline{1-8}
{\bf With random effects} & & & &  -18804.0 & -18802.4 & 0.638 & 0.433 \\  
$\bm\Gamma$    & 0.989   & 0.999   & 0.987     \\
$\mathcal{G}$ & 0.930   & 0.861 &  1.000      \\
\\
{\bf Without random effects} & & & &   -18932.1 & -18935.8 & 2.855 & 2.892\\ 
$\bm\Gamma$    & 0.988   & 0.993   & 0.987     \\
$\mathcal{G}$ & 0.845   & 0.752 &  0.939    \\
\toprule
\end{tabular}}
\end{adjustbox}
\end{table}



 
\section{Analysis of the pharmacogenomic screen}\label{sec:GDSC}

\subsection{Pharmacogenomic data}

We apply our approach to the Genomics of Drug Sensitivity in Cancer (GDSC) database \citep{Yang2013, Garnett2012} to study the relationships between multiple cancer drugs and high-dimensional genomic features characterising cancer cell lines. The pharmacological and genomic data are from the archived dataset release 5 (\url{https://www.cancerrxgene.org}) preprocessed by \cite{Garnett2012}. We would like to investigate how the MRF prior can help to improve inference for groups of drugs that are known to have correlated response; we therefore select two groups of cancer drugs with similar molecular targets and the generic non-targeted chemotherapy agent Methotrexate: four MAPK inhibitors (RDEA119, PD-0325901, CI-1040 AZD6244), two Bcr-Abl tyrosine kinase inhibitors (Nilotinib, Axitinib), and one chemotherapy agent (Methotrexate).

The seven drugs were tested on 499 cell lines from 13 cancer tissue types with complete drug sensitivity values. The drug sensitivity of the cell lines was summarized by the $\log_{10}(\text{IC}_{50})$ from {\it in vitro} drug concentration response experiments. Note that smaller $\log_{10}(\text{IC}_{50})$ values indicate higher sensitivity of a cell line to the drug; therefore a negative regression coefficient indicates that a positive increment of the value of a feature is associated with an increase in drug sensitivity.
In order to explore the relationships between the three groups of drugs and the genomic profiles of the cell lines, we first preselect known cancer genes and their corresponding genomic features by following \cite{Garnett2012}, including 426 copy number variation features (counts) and 68 mutated features (binary). To make a trade-off between the computational efficiency and amount of information from gene expression data, we then preselect three subsets of gene expression features with the largest variances over cell lines, which explain 10\%, 30\% and 50\% of the variation, which results in 269, 1175 and 2602 gene expression features, respectively. This creates three data sets including both gene expression, copy number variation and mutation information, to predict drug sensitivity responses, i.e. feature set I with 763 predictors, feature set II with 1669 predictors, feature set III with 3096 predictors.

\begin{figure}[H]
\centering
\includegraphics[height=0.3 \textwidth]{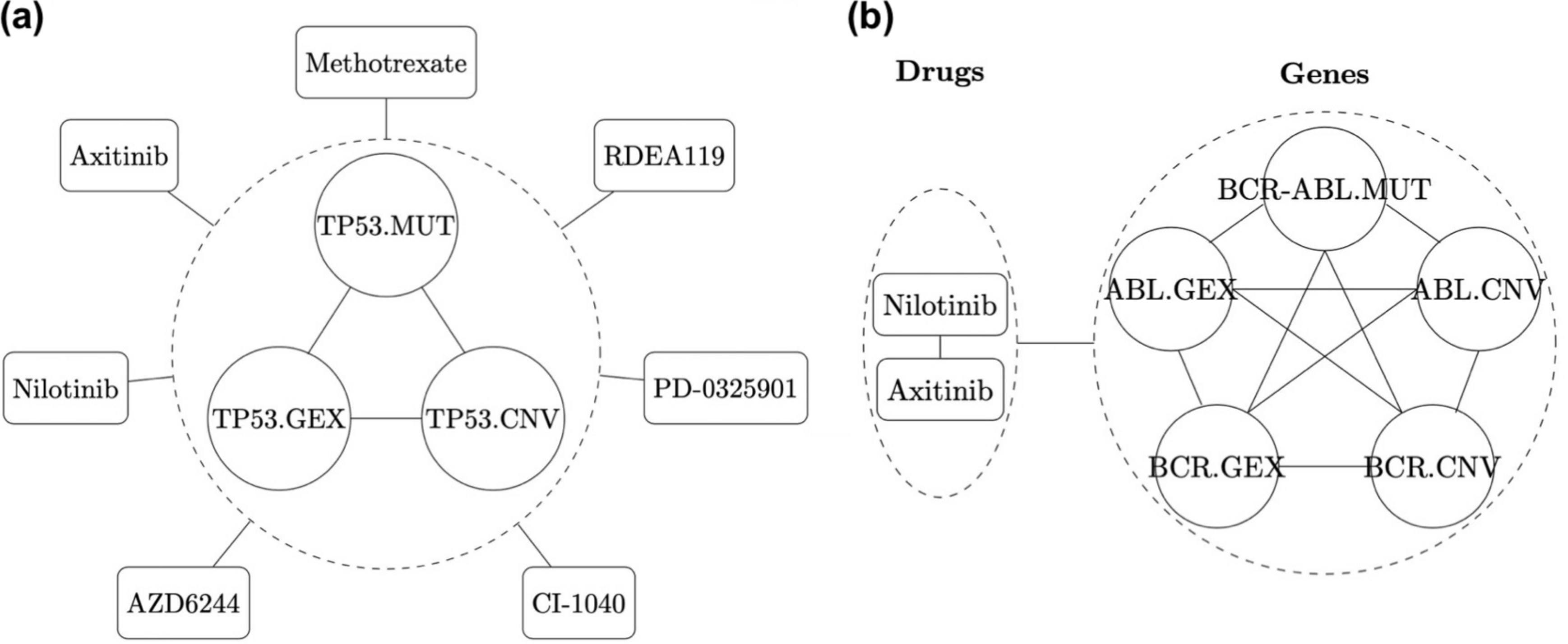}
\caption{GDSC data application: Illustration of the assumed relationships between drugs and related gene features, which are used for the MRF prior. The right panel is for the Bcr-Abl fusion gene, its corresponding related features and the two Bcr-Abl tyrosine kinase inhibitor drugs. The left panel illustrates gene TP35 as one example with its corresponding features representing all three data sources; all seven drugs are shown to indicate that the relationship between the three gene features is valid in relation to all drugs in the data set. The rectangles indicate drugs, solid circles indicate gene features and dashed circles indicate that the elements inside are related. The edges between drugs and dashed circled gene features indicate assumed associations between the gene features and the drug sensitivity measurements for the drugs. The names with suffix ``.GEX", ``.CNV" and ``.MUT" indicate features of expression, copy number variation and mutation, respectively.
}
\label{figure:DrugGenePathway}
\end{figure}

\subsection{Prior specification and model setup}

To construct edge potentials for the MRF prior in the proposed model (\ref{formula:SSUR_RE}) in Section \ref{model:RE}, we summarize some known biological relationships between the drugs and genomic information. First, all features (gene expression, copy number variation, mutation) corresponding to the same gene are assumed to be related. Such group of features are likely to be identified together corresponding to each drug, i.e. if one feature for a certain gene is a predictor of drug sensitivity, then the other features corresponding to the same gene are more likely to be predictors as well. This is illustrated in Figure \ref{figure:DrugGenePathway}(a) and results in a Kronecker product for the edge potentials
\[\scriptsize
\underbrace{G_y}_{7 \text{ drugs}} \otimes \underbrace{G_x}_{3 \text{ features}} - \ \mathbbm{I}_{21} = 
\mathbbm{I}_{7} \otimes 
\begin{pmatrix}
1 & 1 & 1  \\
1 & 1 & 1  \\
1 & 1 & 1 
\end{pmatrix}
- \mathbbm{I}_{21} .
\]
Second, the two Bcr-Abl tyrosine kinase inhibitors were developed to inhibit Bcr-Abl tyrosine kinase activity and proliferation of Bcr-Abl expressing cells, so the point mutation BCR-ABL, and features associated with genes BCR and ABL are related and likely to be identified together corresponding to the two Bcr-Abl inhibitors. This is illustrated in Figure \ref{figure:DrugGenePathway}(b) and results in a Kronecker product for the edge potentials
\[\scriptsize
\underbrace{G_y}_{2 \text{ drugs}} \otimes \underbrace{G_x}_{5 \text{ features}} - \ \mathbbm{I}_{10} = 
\begin{pmatrix}
1 & 1\\
1 & 1
\end{pmatrix}
\otimes 
\begin{pmatrix}
1 & 1 & 1 & 1 & 1 \\
1 & 1 & 1 & 1 & 1 \\
1 & 1 & 1 & 1 & 1 \\
1 & 1 & 1 & 1 & 1 \\
1 & 1 & 1 & 1 & 1  
\end{pmatrix}
- \mathbbm{I}_{10} .
\]
Third, the four MAPK inhibitors were developed to reduce the activity of the MAPK pathway, so genes representing the MAPK pathway are likely to be identified together as potential predictor variables for drug sensitivity of the four MAPK inhibitors. Based on the set of genes linked to the MAPK pathway from the Kyoto Encyclopedia of Genes and Genomes (KEGG) PATHWAY database, we can construct another Kronecker product for the edge potentials. 
Here an edge potential of the $G$ matrix, i.e. an edge weight, is 2, if the corresponding two features are both from the same gene and also belong to one group of drug target genes. Finally, we aggregate the individual Kronecker products by aligning their coordinates in the final $G$ matrix for the MRF prior.

Other prior specifications, MCMC settings and diagnostics can be found in Supplementary S7 and S8. For comparison, we also run almost the same model as SSUR-MRF but with hyperparameter $e=0$ in the MRF prior, which degenerates to a Bernoulli prior, and which we name SSUR-Ber. We choose SSUR-Ber instead of SSUR-hotspot as the comparison model, since it is easier to use than SSUR-hotspot, which has many tuning hyperparameters of the hotspot prior, and because SSUR-Ber has shown similar model performance as SSUR-hotspot shown in Section \ref{sec:simulation}.

\subsection{Results and discussion}

Figure \ref{figure:Gy-MRF} shows an estimated residual structure between the seven drugs by our SSUR-MRF model based on Feature set III with the most genomic information. 
It does not only estimate residual correlation between any two MAPK inhibitors and between the two Bcr-Abl inhibitors, but also separates the chemotherapy drug Methotrexate from the other drugs. Supplementary S9 shows the residual structures between the seven drugs as estimated by the SSUR-MRF and SSUR-Ber models with feature sets I-III, respectively. We find that the structure estimated by our SSUR-MRF model based on feature set III is closest to our knowledge about the relationships between the seven drugs.

\begin{figure}[H]
\centering
\includegraphics[height=0.3 \textwidth]{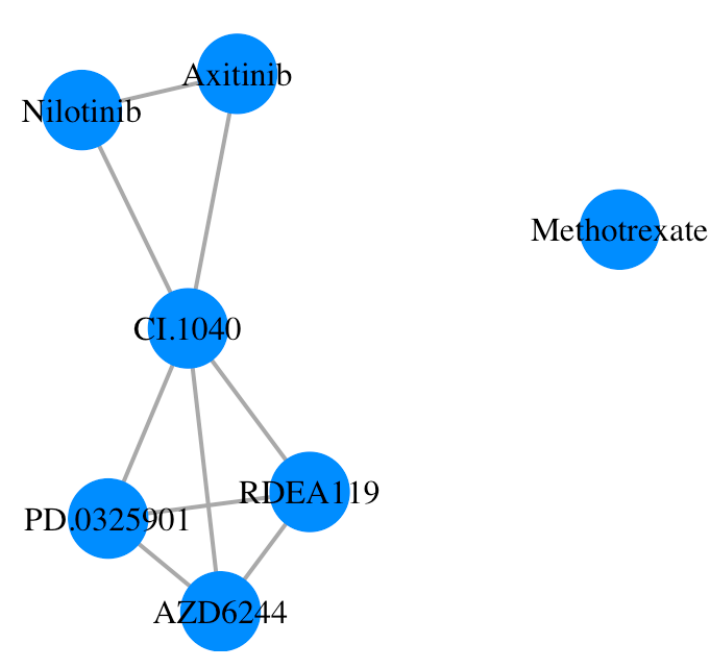}
\caption{GDSC data application: Estimated residual structure between the seven drugs by the SSUR-MRF model based on features set III with $\hat{\mathcal{G}}$ thresholded at 0.5.} 
\label{figure:Gy-MRF}
\end{figure}

To look at variable selection, a gene feature is considered to be identified if the estimated marginal selection probability of its coefficient is larger than 0.5, i.e. if the corresponding latent indicator variable has posterior mean larger than 0.5. Table \ref{tab:gdsc-num-each} reports the numbers of identified features over the seven drugs by the SSUR-Ber and SSUR-MRF models. SSUR-Ber results in very sparse models and identifies a similar number of genomic features for each drug. In contrast, our SSUR-MRF model identifies more genomic features and finds a different model sparsity for the three drug groups, in particular relatively denser models for the four MAPK inhibitors. This indicates that our model is able to distinguish variable selection corresponding to different response variables. For the group with the two BCR-ABL inhibitors, i.e. Nilotinib and Axitinib, both SSUR-Ber and SSUR-MRF identify the mutation BCR-ABL associated with drug Nilotinib, as expected. 

\begin{table}[H]
\centering
\caption{GDSC data application: Number of identified genomic features corresponding to each drug by the SSUR-Ber and SSUR-MRF models.} 
\label{tab:gdsc-num-each}
{
\begin{tabular}{lccccccc}
\toprule
& Nilotinib & Axitinib & RDEA119 & PD-0325901 & CI-1040 & AZD6244 & Methotrexate  \\ \cline{1-8}
{\bf SSUR-Ber }
\smallskip\\
\ \ \ \ Feature set I & 5	& 5	& 2	& 3	& 1	& 0	& 3
\smallskip\\
\ \ \ \ Feature set II  & 1	& 2	& 3	& 1	& 1	& 2	& 2
\smallskip\\
\ \ \ \ Feature set III & 8	& 11	& 8	& 4	& 8	& 10	& 8
\medskip\\
{\bf SSUR-MRF }
\smallskip\\
\ \ \ \ Feature set I  & 1	& 2	& 42	& 41	& 40	& 40	& 0
\smallskip\\
\ \ \ \ Feature set II & 9	& 10	& 56	& 56	& 56	& 57	& 9
\smallskip\\
\ \ \ \ Feature set III & 39	& 38	& 87	& 86	& 86	& 89	& 41 \\
\toprule
\end{tabular}
}
\end{table}

\begin{figure}[H]
\centering
\makebox[\textwidth][c]{\includegraphics[height=0.3 \textwidth]{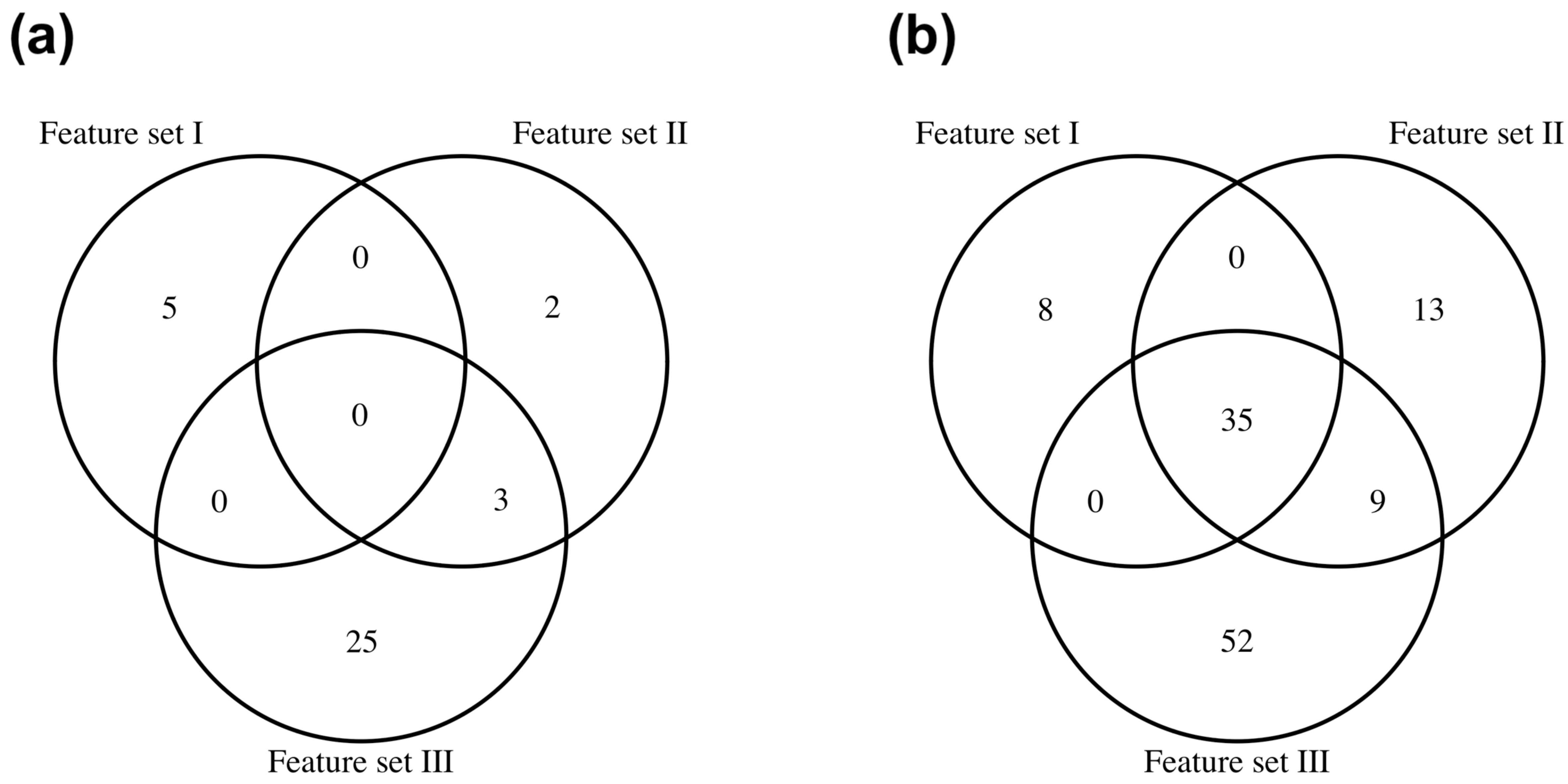}}
\caption{GDSC data application: A Venn diagram for the numbers of identified features for the MAPK inhibitors by SSUR-Ber (panel (a)) and SSUR-MRF (panel (b)) models and overlaps between the models fitted with feature sets I, II, and III.}
\label{figure:venn}
\end{figure}

For the group of the four MAPK inhibitors, Figure \ref{figure:venn} displays the numbers of identified features by SSUR-Ber and SSUR-MRF. For feature sets I, II and III, SSUR-Ber identifies quite different features (Figure \ref{figure:venn}(a)), i.e. there is not much overlap. However, our SSUR-MRF model identifies 35 common features over the three feature sets. This reflects more stable variable selection due to using prior knowledge via the MRF prior. Table \ref{tab:gdsc-num} further shows that the SSUR-Ber model does not identify any known target features for the MAPK inhibitors. Supplementary Table S10.1 shows the identified feature names for the MAPK inhibitors by SSUR-Ber. Overall, our SSUR-MRF model is able to identify many more features than SSUR-Ber, and identifies more known target features for the MAPK inhibitors. 

\begin{table}[H]
\centering
\caption{GDSC data application: Numbers of genomic features selected as predictors for the MAPK inhibitors in the SSUR-Ber and SSUR-MRF models. Note that Feature sets I, II and III include 40, 55, and 81 features corresponding to known target genes of the corresponding drug, respectively.} 
\label{tab:gdsc-num}
{
\begin{tabular}{lcc}
\toprule
\smallskip
 & SSUR-Ber (\%) & SSUR-MRF (\%)  \\ \cline{1-3}
{\bf Feature set I}
\smallskip\\
\ \ \ \ $\frac{\text{\# identified targets}}{\text{\# known targets}}$ & 0/40 (0\%) & 7/40 (17.5\%)
\smallskip\\
\ \ \ \ $\frac{\text{\# identified targets}}{\text{\# identified features}}$ & 0/5 (0\%) & 7/43 (16.3\%) 
\smallskip\\
{\bf Feature set II} 
\smallskip\\
\ \ \ \ $\frac{\text{\# identified targets}}{\text{\# known targets}}$ & 0/55 (0\%) & 8/55 (14.5\%)
\smallskip\\
\ \ \ \ $\frac{\text{\# identified targets}}{\text{\# identified features}}$  & 0/5 (0\%) & 8/57 (14.0\%)  
\smallskip\\
{\bf Feature set III} 
\smallskip\\
\ \ \ \ $\frac{\text{\# identified targets}}{\text{\# known targets}}$ & 0/81 (0\%) & 17/81 (21.0\%)
\smallskip\\
\ \ \ \ $\frac{\text{\# identified targets}}{\text{\# identified features}}$  & 0/28 (0\%) & 17/96 (17.7\%) \\
\toprule
\end{tabular}
}
\end{table}

\begin{figure}[H]
\centering
\makebox[\textwidth][c]{\includegraphics[height=0.3 \textwidth]{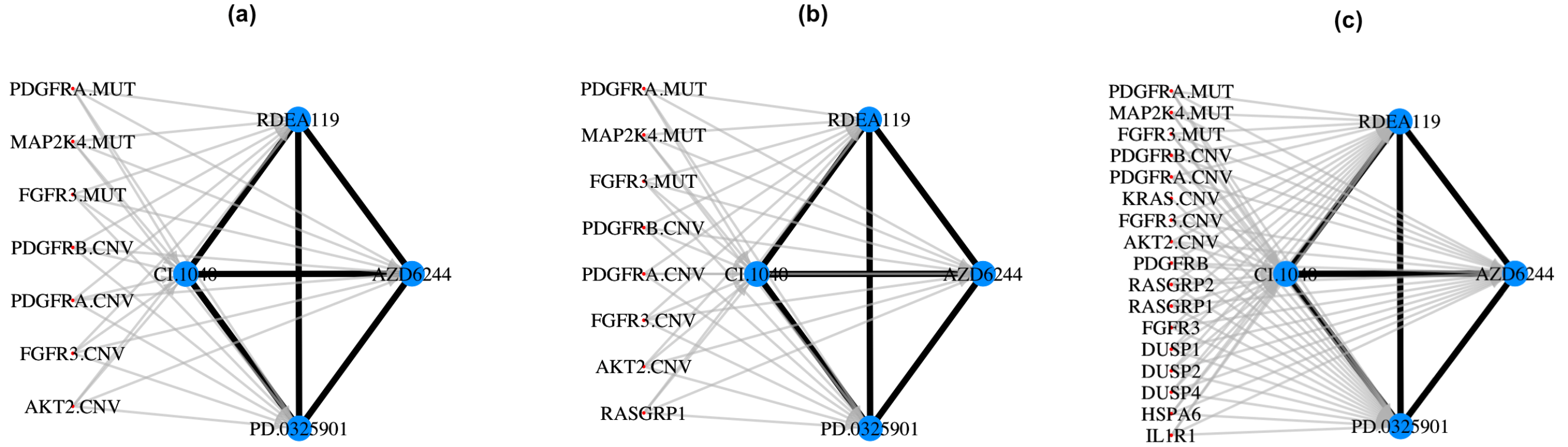}}
\caption{GDSC data application: Estimated network between the MAPK inhibitors and identified target genes based on $\hat{\mathcal{G}}$ and $\hat{\bm{\Gamma}}$ thresholded at 0.5 by SSUR-MRF corresponding to feature set I, II and III respectively.}
\label{figure:Gy-all}
\end{figure}

Figure \ref{figure:Gy-all} shows the names of features that were identified for the MAPK inhibitors by the SSUR-MRF model. The seven copy number variation and mutation features in Figure \ref{figure:Gy-all}(a) are also in Figure \ref{figure:Gy-all}(b) and (c), because only more gene expression features are selected by the models using feature sets II and III, but no additional mutation or copy number variation features. As more target gene expression features are used to construct the edge potentials in the MRF prior in the models built with feature sets II and III, our approach can identify more of them. As Figure \ref{figure:venn}(b) shows, we have identified 35 common features with the SSUR-MRF model over feature sets I, II and III, but only seven of these common features belong to known target genes of the corresponding drugs as shown in Figure \ref{figure:Gy-all}. We find that the 28 other common identified features (listed in Supplementary Table S10.2) are cancer genes, i.e. genes that are known to be deregulated in cancer. The Cancer Gene Census summarizes how dysfunction of these genes drives cancer \citep{Sondka2018}. 

In Table \ref{tab:gdsc-prediction}, prediction performances of the SSUR-Ber and SSUR-MRF models are reported based on Feature set III which has the most genomic information. Overall, prediction performance is very similar between the two models. As for $\widehat{\text{elpd}}_{\text{loo}}$ or $\widehat{\text{elpd}}_{\text{waic}}$, our SSUR-MRF model is slightly worse than SSUR-Ber. To assess the prediction performance of the median probability model, we need an independent data set for out-of-sample prediction to obtain RMSPE. For this purpose we gathered 46 cell lines with complete pharmacogenomic data from the updated GDSC data set by \cite{Smirnov2016}, that were not included in our training data. Table \ref{tab:gdsc-prediction} shows that SSUR-MRF has slightly better RMSPE than SSUR-Ber on this independent data set.

\begin{table}[H]
\centering
\caption{GDSC data application: Prediction performance of the SSUR-Ber and SSUR-MRF models based on Feature set III.} 
\label{tab:gdsc-prediction}
{
\begin{tabular}{rcc}
\toprule
\smallskip
 & SSUR-Ber & SSUR-MRF \\ \cline{1-3}
elpd.LOO & -8135.5 & -8143.1 \\
elpd.WAIC & -8168.8 & -8178.0 \\
RMSE & 2.003 & 1.883 \\
RMSPE & 2.095 & 2.062 \\
\toprule
\end{tabular}
}
\end{table}




\begin{figure}[H]
\centering
\includegraphics[height=1 \textwidth]{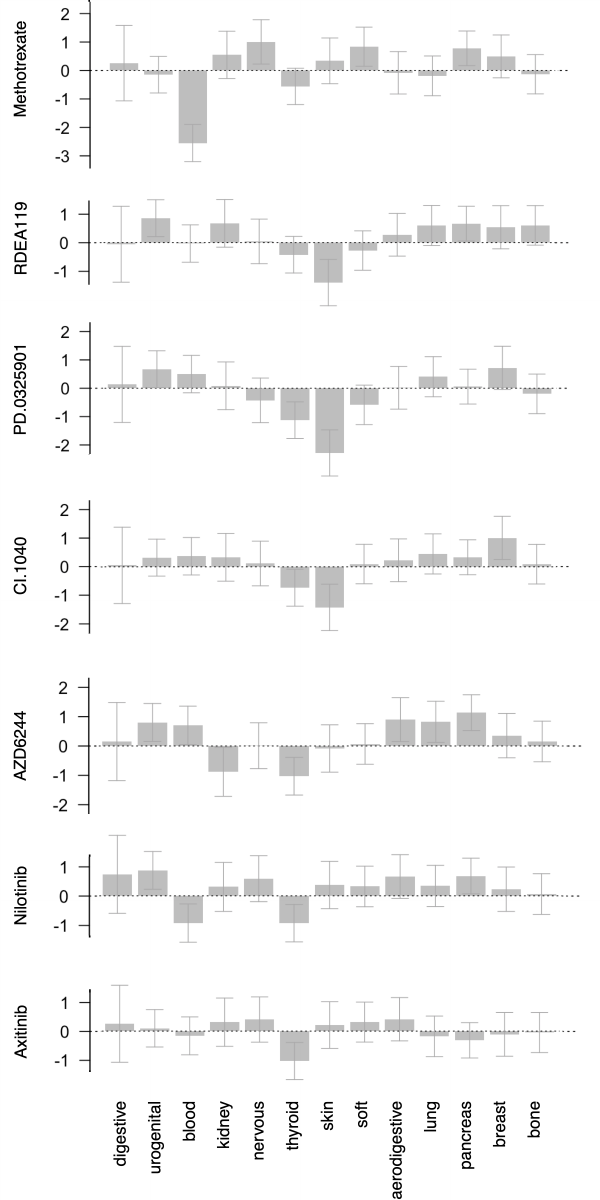}
\caption{GDSC data application: Posterior estimates for the cancer tissue random effects for all drugs based on the median probability model. Random effects are centred around zero. Error bars are $\pm$ standard deviation of the posterior mean.}
\label{GDSC_effects}
\end{figure}

Our approach also estimates the tissue-specific effects of 13 cancer types, which may indicate relationships between drug responses and cancer types. Figure \ref{GDSC_effects} shows the estimated random effects by SSUR-MRF using the genomic feature set III. Negative effect estimates can indicate especially high effectiveness of a drug to kill cancer cells of the corresponding cancer type. We focus on the strongest (negative) effects, since most error bars of the posterior mean for the cancer type random effects are quite large. Methotrexate has the strongest average effect in blood cancer samples; it is known to be an effective chemotherapeutic agent in leukemia \citep{Powell2010}. Supplementary S12 shows that Methotrexate has much lower $\log$(IC$_{50}$) values (i.e. more effectiveness) on cell lines from blood tissue type compared with other tissue types. Three of the four MAPK inhibitors (RDEA119, PD-0325901, CI-1040) have their strongest effect in skin cancer cell lines. Supplementary S12 also shows that these drugs have lower $\log$(IC$_{50}$) values on skin tissue cell lines compared with other tissue types, while AZD6244 shows more variation. Nilotinib and Axitinib are common targeted therapies for chronic myelogenous leukemia with a BCR-ABL mutation \citep{Halbach2016}. We can observe the effect of Nilotinib on blood cancer samples in Figure \ref{GDSC_effects} although with large error bars, and the quite low $\log$(IC$_{50}$) on the only four BCR-ABL mutated blood cancer cell lines are shown in Supplementary S12.


 
\section{Conclusion}\label{sec:conclusion} 
\subsection*{}

In this work, we have developed a multivariate Bayesian structured variable selection model for analyzing data from pharmacogenomic studies. Our model exploits the relationships between multiple correlated response variables (drug sensitivity measurements) and high-dimensional structured multi-omics input data for variable selection and to improve prediction. With our approach we want to {\bf (a)} be able to borrow known information between response variables and predictors, {\bf (b)} learn associations between response variables and predictors, and {\bf (c)} understand the residual covariance of response variables. The proposed approach allows us to make use of known network information on the relationships between responses and predictors in an MRF prior for the variable selection indicator $\bm{\Gamma}$, and to further simultaneously select predictors in a sparse manner and learn the residual covariance matrix between the response variables. In addition, we can take into account sample heterogeneity through random effects which are excluded from the variable selection. Guidance for specifying (weakly) informative hyper-parameters has been provided in the Supplement. 

Through the simulation studies, we have demonstrated that the proposed approach can recover the network structure (i.e. latent indicator variable $\bm\Gamma$) between multiple response variables and predictors, and predict responses well. We have found that including only a small amount of prior knowledge for most patterns/network groups (Figure \ref{fig:sensitivityII}(d)) will improve model performance over a model that does not any include prior knowledge. 
Our approach is also robust to noise in the prior information (i.e. false edge potentials) in the MRF prior (Table \ref{tab:sensitivity}). Even if there is no prior association knowledge between drugs and genes/pathways (i.e. subgraphs), our approach has similar model performance as SSUR-hotspot.

In the pharmacogenomic data application, our approach robustly identified molecular targets of the targeted therapies, and also validated other known cancer-related genes. The use of known information in the MRF prior improved the prediction performance in the independent validation data compared to SSUR-Ber when applied to the largest input data set (feature set III). Through the random effects in our approach, cancer tissue effects were estimated, which could indicate potential relationships between drugs and cancer types. Nevertheless, there was still remaining heterogeneity within cancer types, e.g. reflecting molecular cancer sub-types. To address this, our model could be extended to multilevel random effects or a mixture approach could be employed for the random effects, e.g. by a flexible Dirichlet process prior \citep{LiLin2010, Heinzl2012}.

Although our approach has been successfully applied in scenarios with multiple correlated response variables and high-dimensional predictors, it might become too computationally demanding if the model is not assumed to be very sparse (i.e. if the number of associated features is not assumed to be much smaller than $mp$). An alternative is to change our MCMC sampling approach to approximate inference, e.g. variational inference \citep{Blei2017, Zhang2019, Munch2021}.


\section*{Supplementary Material}
R package {\bf BayesSUR} is available on the Comprehensive R Archive Network at \url{https://CRAN.R-project.org/package=BayesSUR}. R code reproducing all results can be found at \url{https://github.com/zhizuio/BayesSUR-RE}. The Supplementary material contains details on prior specifications, sampling steps, calculation of model performance indexes, details on the simulation scenarios and sensitivity analysis, and additional results of the GDSC data analysis.


\section*{Acknowledgements}
This work was supported by 
Research Council of Norway project No. 237718 ``Big Insight'' (ZZ), European Union Horizon 2020 grant agreements No. 847912 ``RESCUER'' (MZ) and No. 633595 ``DynaHealth'' (AL), UK Medical Research Council grants MR/M013138/1 (MB, AL). The authors thank Dr. Leonardo Bottolo for discussions. 

\bibliographystyle{apalike} 
\bibliography{refs}

\clearpage

\setcounter{page}{1}
\pagenumbering{roman}
\captionsetup[table]{labelformat=empty}

{\bf\Large Supplementary materials for `Multivariate Bayesian structured variable selection for pharmacogenomic studies'}
\medskip

\section*{S1: Prior specifications for $\mathcal{HIW}$ prior and spike-and-slab prior}
\subsection*{}

For the hyper-parameters $\nu$ and $\tau$ in $\Psi \sim  \mathcal{HIW}_{ \mathcal{G}}(\nu, \tau \mathbb{I}_m)$, we specify a fixed $\nu=m+2$ and $\tau \sim  \mathcal{G}amma(a_{\tau}, b_{\tau})$. $\nu=m+2$ is suggested by \citet{Bottolo2021}, because it is the smallest integer degree resulting in a proper prior for the reparametrized parameter $\sigma_j^2$. A sensitivity analysis for these hyper-parameters of the hyper-inverse Wishart prior can be found in Supplementary S5. 

The hyper-parameters $a_w$ and $b_w$ control the variance of the spike-and-slab prior for non-zero regression coefficients in \begin{equation*}
\beta_{kj}| \gamma_{kj},w \sim \gamma_{kj} \mathcal{N}(0,\ w) + (1-\gamma_{kj})\delta_0(\beta_{kj}),\\
\end{equation*} 
which determines the posterior scale for the estimated effects. In order to provide a sufficiently large scale for the effects, we would like to allow some posterior density for values of $w$ that will correspond to non-neglible posterior density for values of $\beta_{kj}$ larger than the upper 95\% confidence bound $\mathbb{E}[\beta_{kj}] + 1.96 \sqrt{\mathbb{V}\text{ar}[\beta_{kj}]}$, where $\mathbb{E}[\cdot]$ and $\mathbb{V}\text{ar}[\cdot]$ are the mean and variance of the prior for the effect $\beta_{kj}$. This is to ensure that prior of $w$ provides large enough variation to be able to cover a wide range of $\beta_{kj}$. Since $w$ has posterior conditional
\begin{equation*}
w| a_{w},b_{w},\bm{\Gamma},\bm{B} \sim  \mathcal{IG}\left( a_{w}+\frac{1}{2}\sum_{kj}\gamma_{kj}, b_{w}+\frac{1}{2}\sum_{k,j}\beta_{kj}^2 \right),\tag{$\star$}
\label{formula:postW}
\end{equation*}
we can choose proper $a_w$ and $b_w$ according to this posterior conditional with 5\% quantile larger than $\mathbb{E}[\beta_{kj}] + 1.96 \sqrt{\mathbb{V}\text{ar}[\beta_{kj}]}$. Since $a_w$ and $b_w$ are combined with the factors $\frac{1}{2}\sum_{kj}\gamma_{kj}$ and $\frac{1}{2}\sum_{kj}\beta_{0,tj}^2$ respectively, $a_w$ and $b_w$ can be chosen with similar scales as the two factors.


For example, we assume a model sparsity (i.e., proportion of nonzero coefficients) $r_{sparsity}=\frac{1}{mp}\sum_{kj}\gamma_{kj} \in (0,1)$, and $\mathbb{E}[\beta_{kj}]$ and $\mathbb{V}\text{ar}[\beta_{kj}]$ can either be estimated roughly from previous studies or elucidated using expert knowledge about typical effect sizes in similar studies. Then we specify
\begin{align*}
a_w = const_a \cdot 1/2 \cdot m \cdot p \cdot r_{sparsity} ,
b_w = const_b \cdot 1/2 \cdot m \cdot p \cdot r_{sparsity} \cdot (\mathbb{E}[\beta_{kj}])^2,
\end{align*}
where $const_a$ and $const_b$ are chosen to ensure that the 5\% quantile of the posterior conditional ($\star$) is larger than $\mathbb{E}[\beta_{kj}] + 1.96 \sqrt{\mathbb{V}\text{ar}[\beta_{kj}]}$. 

\section*{S2: Prior specification for the MRF prior}
\subsection*{}

It is known that the hyper-parameter $e$ in a Markov random field (MRF) prior
$$f(\bm{\gamma} | d, e, G) \propto \exp \left\{d\mathbbm{1}^\top\bm{\gamma} +e\bm{\gamma}^\top G \bm{\gamma} \right\} $$ 
can display a phase transition behaviour \citep{Li2010, Stingo2011} which results in a sharp increase in the number of selected predictors (i.e., $\beta_{kj} \ne 0$) with a small change of $e$ given a value $(w, d)$. In the MRF prior, the model sparsity is logit$^{-1}d$ which can be used to specify the hyper-parameter $d$. Here logit is log-odds that is the natural logarithm of the odds. 

To specify $e$, \citet{Stingo2011, Lee2017} first looked for the phase transition boundary, and then implemented a grid search for the hyper-parameter. This might still search many values for $e$ over the phase transition boundary, which would result in very dense models in high-dimensional settings. So we first determine the expected largest value $e$ to avoid searching in the space of too dense models which would be computationally slow, and then implement the grid search strategy. If one wants to specify model sparsity in the range $(c_1,c_2)$, one can achieve this by specifying the hyper-parameter $d$ subject to logit$^{-1}d=c_1$ and search hyper-parameter $e$ from 0 to $-d - \ln(c_2^{-1} -1)$ (see below). Because of $d = $logit $c_1$, $c_1$ represents a lower bound for the model sparsity which is reached if e = 0. We aim to give a roughly estimate for $e\ge0$ based on the assumed largest sparsity $c_2$.
\medskip\\
\noindent{\bf An estimate for $e$ in the MRF prior}
\subsection*{}

A MRF prior 
$$f(\bm{\gamma} | d, e, G) \propto \exp \left\{d\mathbbm{1}^\top\bm{\gamma} +e\bm{\gamma}^\top G \bm{\gamma} \right\} $$ 
corresponds to a set of conditional Bernoulli distributions given by
$$f(\gamma_j | \bm{\gamma}_{-(j)} d, e, G) = p_j^{\gamma_j}(1-p_j)^{1-{\gamma_j}},$$
where
$$p_j = \frac{1}{1+\exp(-d - e\sum_{r \ne j}g_{rj}\gamma_r)},$$
and $\{g_{rj}\}_{rj} = G$.

One might expect to take the average over $\bm{\gamma}$ to be the assumed largest sparsity $c_2$ (i.e., roughly inclusion probability of each predictor),
$$\frac{1}{mp}\sum_{j=1}^{mp} \frac{1}{1+\exp(-d - e\sum_{r \ne j}g_{rj}\gamma_r)} = c_2.$$
However, it is not easy to get an explicit or approximate expression for $e$ via the equation above.

Our strategy is to assume at least one link/neighbour so that the parameter $e$ is not too small. Since $\sum_{r \ne j}g_{rj}\gamma_r$ counts the number of links/neighbours of the $r$th feature, we let 
$$\frac{1}{1+\exp(-d - e\cdot 1)} = c_2$$
$$\Rightarrow e = -d - \ln(c_2^{-1} -1).$$

\clearpage
\section*{S3: Sampling steps for the parameters in SSUR-MRF model}
\subsection*{}

Here the Gibbs samplers use the posterior conditionals of the parameters referring to \cite{Bottolo2021}.

\begin{itemize}
\item sampling latent indicator variables $\bm{\Gamma}$ using the Metropolis-Hastings sampler;
\item sampling hyper-parameter $\tau$ using a random walk Metropolis sampler;
\item sampling hyper-parameter $w$ (and $w_0$) using Gibbs sampling;
\item sampling the graph $ \mathcal{G}$ from the junction tree sampler;
\item sampling $\bm{\sigma}^2$ and $\bm{\rho}$ from the full conditional distributions (which would be the Gibbs sampler);
\item sampling coefficients $\bm{B}$ (and $\bm{B}_0$) from the full conditional distributions (which would be the Gibbs sampler).
\end{itemize}

\section*{S4: Algorithms for simulation scenarios}
\subsection*{}

\begin{algorithm}[H]
\caption{Simulation steps without random effects}
\label{alg:protocol1}
\begin{algorithmic}[1]
\STATE Design a decomposable graph $\mathcal{G}$, dim$(\mathcal{G})=m \times m$
\STATE Design a sparse matrix $\bm{\Gamma}$, dim$(\Gamma)=p \times m$
\STATE Simulate $x_{ik} \sim \mathcal{N}(0,1)$ and $x^*_{ik} \sim \mathcal{N}(0,1)$, $i=1,\cdots,n$ and $k=1,\cdots,p$
\STATE Simulate $\alpha_j$, $\beta_{kj} \sim \mathcal{N}(0,1)$, $k=1,\cdots,p$ and $j=1,\cdots,m$
\STATE Simulate $\tilde{u}_{ij} \sim \mathcal{N}(0,0.5^2)$, $i=1,\cdots,n$ and $j=1,\cdots,m$
\STATE Simulate $P \sim \mathcal{W}_{\mathcal{G}}(3,M)$ where diagonals of $M$ are 1 and off-diagonals are 0.5, dim$(P)=m \times m$ 
\STATE Use Cholesky decomposition $\text{chol}(P^{-1})$ to get $\mathbf{U} = \tilde{\mathbf{U}} \cdot \text{chol}(P^{-1})$ 
\STATE Generate $\mathbf{Y} =  \mathbbm{1}\bm{\alpha}^\top+(\mathbf{X}\bm{B})_{\bm{\Gamma}}+ \mathbf{U}$ and $\mathbf{Y}^* =  \mathbbm{1}\bm{\alpha}^\top+\mathbf{X}^*\bm{B}_{\bm{\Gamma}}+ \mathbf{U}$
\end{algorithmic}
\end{algorithm} 

\begin{algorithm}[H]
\caption{Simulation steps with random effects}
\label{alg:protocol2}
\begin{algorithmic}[2]
\STATE Design a decomposable graph $\mathcal{G}$, dim$(\mathcal{G})=m \times m$
\STATE Design a sparse matrix $\bm{\Gamma}$, dim$(\Gamma)=p \times m$
\STATE Simulate $x_{ik} \sim \mathcal{N}(0,1)$ and $x^*_{ik} \sim \mathcal{N}(0,1)$, $i=1,\cdots,n$ and $k=1,\cdots,p$
\STATE Simulate $\beta_{kj} \sim \mathcal{N}(0,1)$, $k=1,\cdots,p$ and $j=1,\cdots,m$
\STATE Simulate $\bm{z}_{i} \sim multinomial(0.1,0.2,0.3,0.4)$ where $\bm z_i =(z_{i1},\cdots,z_{iT})'$, $i=1,\cdots,n$, $T=4$, and the same for and $\bm{z}^*_{i}$
\STATE Simulate $\beta_{0,t1} \sim \mathcal{N}(0,2^2)$, $t=1,\cdots,T$
\STATE Simulate $\tilde{u}_{ij} \sim \mathcal{N}(0,0.5^2)$, $i=1,\cdots,n$ and $j=1,\cdots,m$
\STATE Simulate $P \sim \mathcal{W}_{\mathcal{G}}(3,M)$ where diagonals of $M$ are 1 and off-diagonals are 0.5, dim$(P)=m \times m$
\STATE Use Cholesky decomposition $\text{chol}(P^{-1})$ to get $\mathbf{U} = \tilde{\mathbf{U}} \cdot \text{chol}(P^{-1})$
\STATE Generate $\mathbf{Y} = \mathbf{X}\bm{B}_{\bm{\Gamma}}+ \mathbf{ZB}_0 + \mathbf{U}$ and $\mathbf{Y}^* = \mathbf{X}^*\bm{B}_{\bm{\Gamma}}+ \mathbf{Z}^*\bm{B}_0 + \mathbf{U}$
\end{algorithmic}
\end{algorithm} 

To specify hyperparameters in Algorithm 1, we set $a_{w}=15$ and $b_{w}=60$ for the variance of shrinkage coefficients by choosing $r_{sparsity}=0.1$, $\mathbb{E}[\beta_{kj}]=-0.1$, $\mathbb{V}\text{ar}[\beta_{kj}]=1$, $const_{a_w}=1/20$ and $const_{b_w}=20$ according to the prior specification in Section S5. To specify the hyperparameter of random effects in Algorithm 2, we set $a_{w_0}=100$ by choosing $r_{sparsity}=1$, $\mathbb{E}[\beta_{0,kj}]=0$, $\mathbb{V}\text{ar}[\beta_{0,kj}]=2^2$, $const_{a_{w_0}}=1/30$ and $b_{w_0}=500$ according to the prior specification in Section S3 by setting $r_{sparsity}=1$ (i.e. no variable selection for the random effects) and $\mathbb{E}[\beta_{0,tj}]=0$ (i.e. random effects centred at zero). For the hyperparameters $d$ and $e$ in a MRF prior, we assume the model sparsity in the range $(c_1,c_2)=(0.1,0.3)$, and then $d=\text{logit } c_1=-2$ and grid interval $e=(0, -d - \ln(c_2^{-1} -1))=(0,1.2)$. We determine the final $e=1$ for Algorithm 1, and the final $e=0.2$ for Algorithm 2.

We run a MCMC sampler with 5 chains, 400 000 iterations in total with the first 200 000 iterations as a burn-in period. 

\section*{S5: Sensitivity analysis the hyper-inverse Wishart prior in SSUR-MRF model} 
\subsection*{}

For our approach SSUR-MRF model, we tested various hyperparameters of the hyper-inverse Wishart prior based on the simulation scenario 1. Large $\nu=100$ or $a_{\tau}$ in the hyper-inverse Wishart prior results in a denser $\hat{\mathcal{G}}$ than $\mathcal{G}$ (Figure S5.1, S5.3, S5.5 and S5.7). But all hyperparameters of the hyper-inverse Wishart prior are not much sensitive to the structure recovery of $\mathbf{\Gamma}$ and response predictions (Table S3).

\begin{figure}[H]
\centering
\includegraphics[height=0.35 \textwidth]{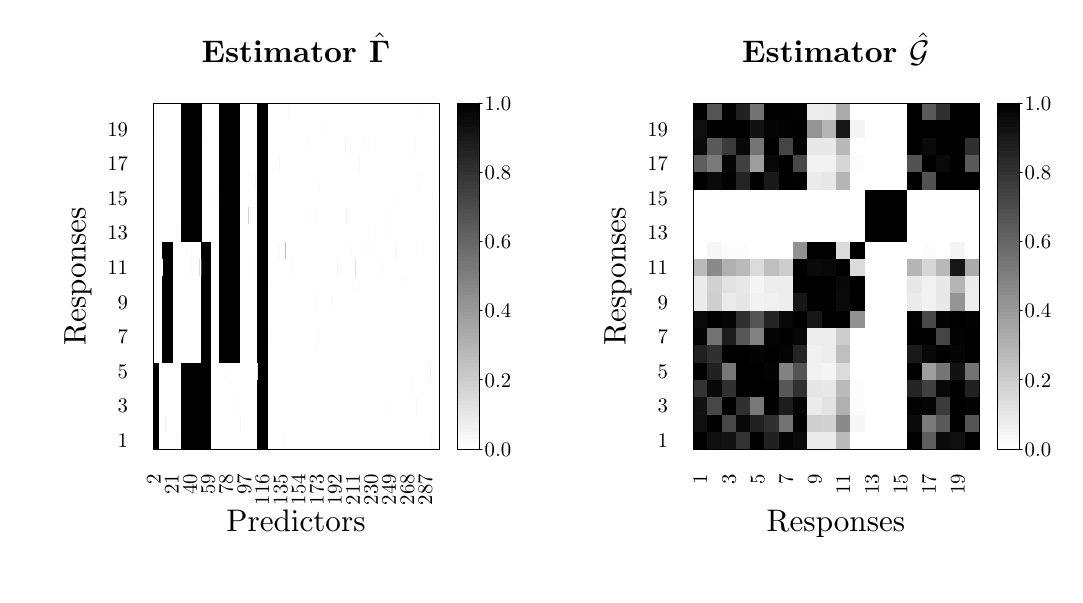}
\caption*{\it Figure S5.1: Posterior mean of the latent indicator variable $\hat{\mathbf{\Gamma}}$ and learning structure $\hat{\mathcal{G}}$ by {\bf SSUR-MRF}, $\mathcal{HIW}_{\mathcal{G}}({\bf 100},\tau \mathbbm{I})$, $\tau \sim \mathcal{G}amma(0.1,10)$ and $\eta \sim \mathcal{B}eta(0.1,1)$.}
\label{fig:supplSensitivityMoreNu}
\end{figure}

\begin{figure}[H]
\centering
\includegraphics[height=0.35 \textwidth]{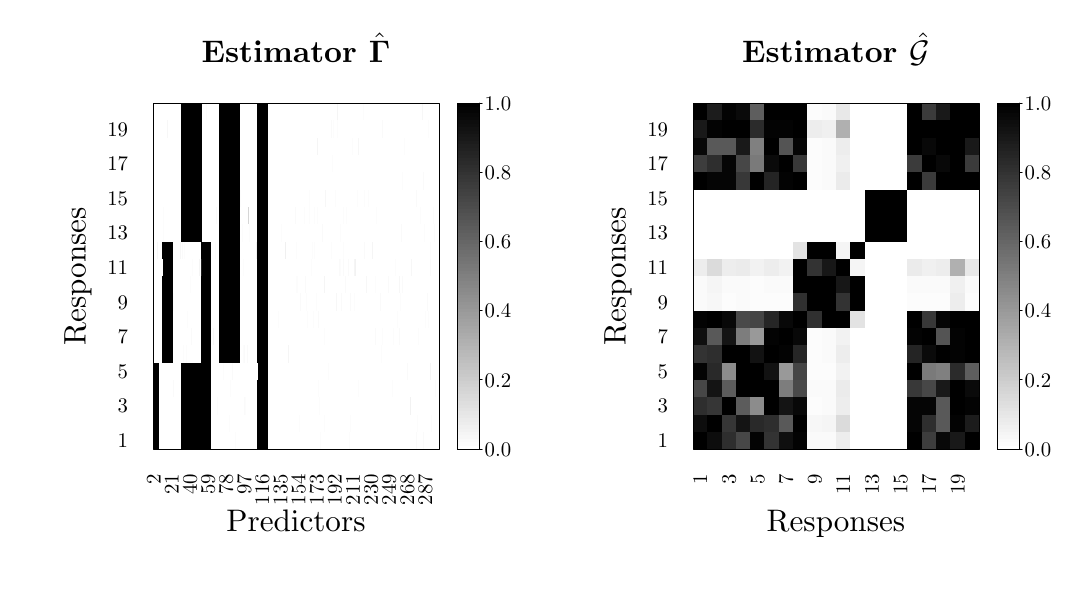}
\caption*{\it Figure S5.2: Posterior mean of the latent indicator variable $\hat{\mathbf{\Gamma}}$ and learning structure $\hat{\mathcal{G}}$ by {\bf SSUR-MRF} $\mathcal{HIW}_{\mathcal{G}}(22,\tau \mathbbm{I})$, $\tau \sim \mathcal{G}amma({\bf 0.001},10)$ and $\eta \sim \mathcal{B}eta(0.1,1)$.}
\label{fig:supplSensitivityMoreAtau-6}
\end{figure}

\begin{figure}[H]
\centering
\includegraphics[height=0.35 \textwidth]{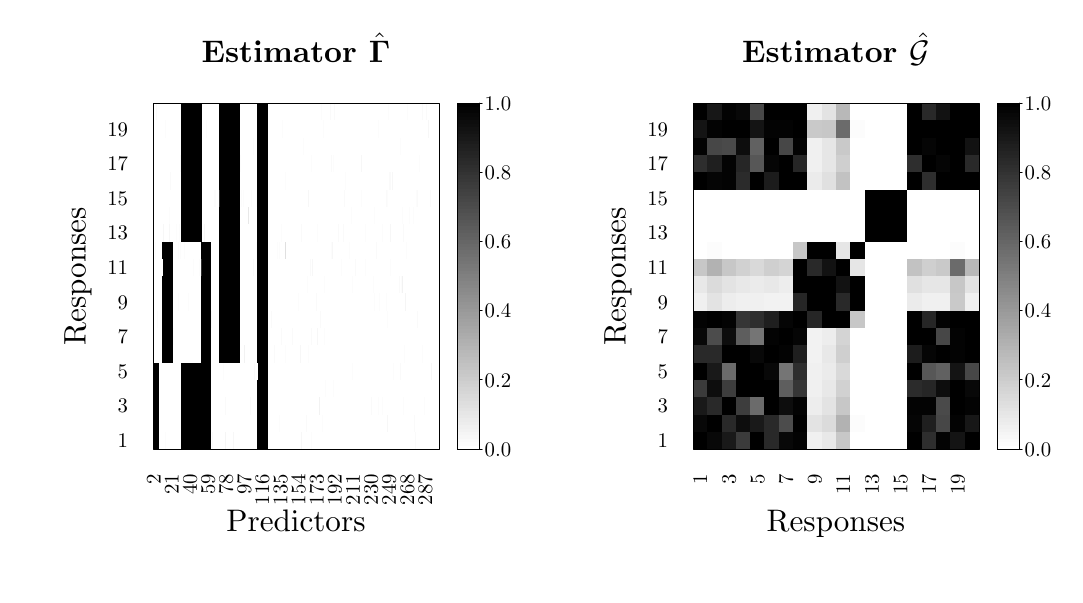}
\caption*{\it Figure S5.3: Posterior mean of the latent indicator variable $\hat{\mathbf{\Gamma}}$ and learning structure $\hat{\mathcal{G}}$ by {\bf SSUR-MRF}, $\mathcal{HIW}_{\mathcal{G}}(22,\tau \mathbbm{I})$, $\tau \sim \mathcal{G}amma({\bf 100},10)$ and $\eta \sim \mathcal{B}eta(0.1,1)$.}
\label{fig:supplSensitivityMoreAtau6}
\end{figure}

\begin{figure}[H]
\centering
\includegraphics[height=0.35 \textwidth]{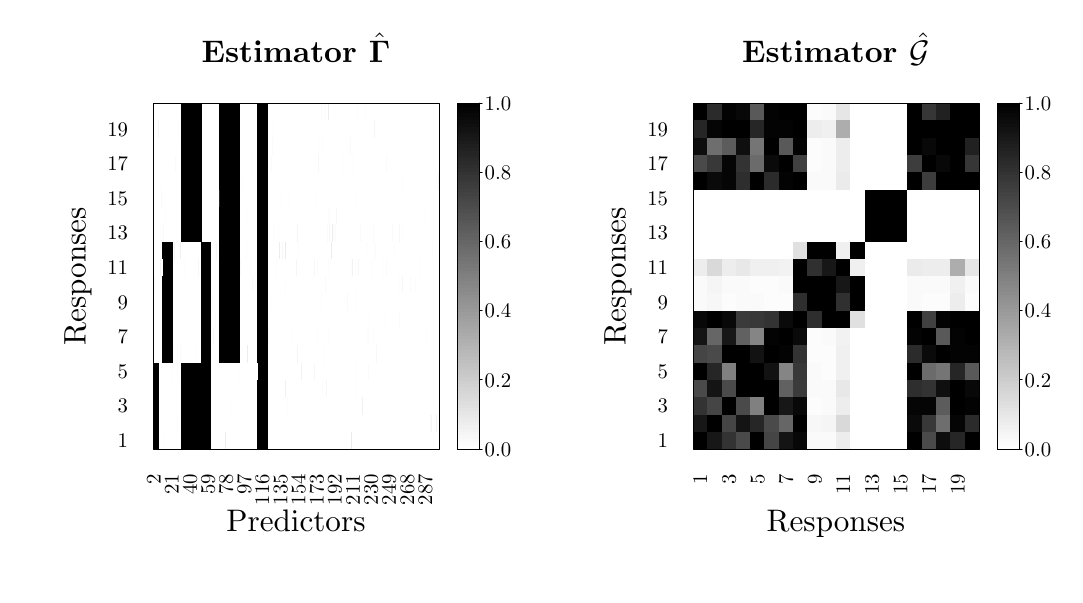}
\caption*{\it Figure S5.4: Posterior mean of the latent indicator variable $\hat{\mathbf{\Gamma}}$ and learning structure $\hat{\mathcal{G}}$ by {\bf SSUR-MRF}, $\mathcal{HIW}_{\mathcal{G}}(22,\tau \mathbbm{I})$, $\tau \sim \mathcal{G}amma(0.1,10)$ and $\eta \sim \mathcal{B}eta({\bf 0.001},1)$.}
\label{fig:supplSensitivityMoreAeta-6}
\end{figure}

\begin{table}[H]
\centering
\caption*{\it  Table S5: Performance of variable selection and response predictions by SSUR-MRF
with different hyper-inverse Wishart prior} 
\label{tab:supplSensitivity}
{\scriptsize
\begin{tabular}{lcccccccl} 
\hline
\hline
\smallskip
& accuracy & sensitivity & specificity & $\widehat{\text{elpd}}_{\text{loo}}$ & $\widehat{\text{elpd}}_{\text{waic}}$ & RMSE & RMSPE \\ \hline
$\nu=22$, $\tau \sim \mathcal{G}amma(0.1,10)$, $\eta \sim \mathcal{B}eta(0.1,1)^{\dag}$  & 0.989 & 0.998 & 0.986 & -18616.6 & -18616.8 & 0.643 & 0.412  \\
$\nu=100$, $\tau \sim \mathcal{G}amma(0.1,10)$, $\eta \sim \mathcal{B}eta(0.1,1)$     & 0.989 & 0.998 & 0.987 & -18656.6 & -18657.2 & 0.646 & 0.411   \\
$\nu=22$, $\tau \sim \mathcal{G}amma(0.001,10)$, $\eta \sim \mathcal{B}eta(0.1,1)$     & 0.989 & 0.998 & 0.987 & -18617.5 & -18617.7 & 0.644 & 0.410   \\
$\nu=22$, $\tau \sim \mathcal{G}amma(100,10)$, $\eta \sim \mathcal{B}eta(0.1,1)$     & 0.989 & 0.998 & 0.987 & -18617.6 & -18617.7 & 0.648 & 0.413   \\
$\nu=22$, $\tau \sim \mathcal{G}amma(0.1,10)$, $\eta \sim \mathcal{B}eta(0.001,1)$     & 0.989 & 0.998 & 0.987 & -18617.5 & -18617.6 & 0.643 & 0.413 \\ 
\hline
\hline
\multicolumn{6}{p{14cm}}{$^{\dag}$These specifications are the same as the SSUR-MRF model in Table 1.} \\
\end{tabular}
}
\end{table}

\clearpage
\section*{S6: Estimated $\mathcal{G}$ for the sensitivity analysis in Section 3.3}
\subsection*{}


\begin{figure}[H]
\centering
\makebox[\textwidth][c]{\includegraphics[height=0.3 \textwidth]{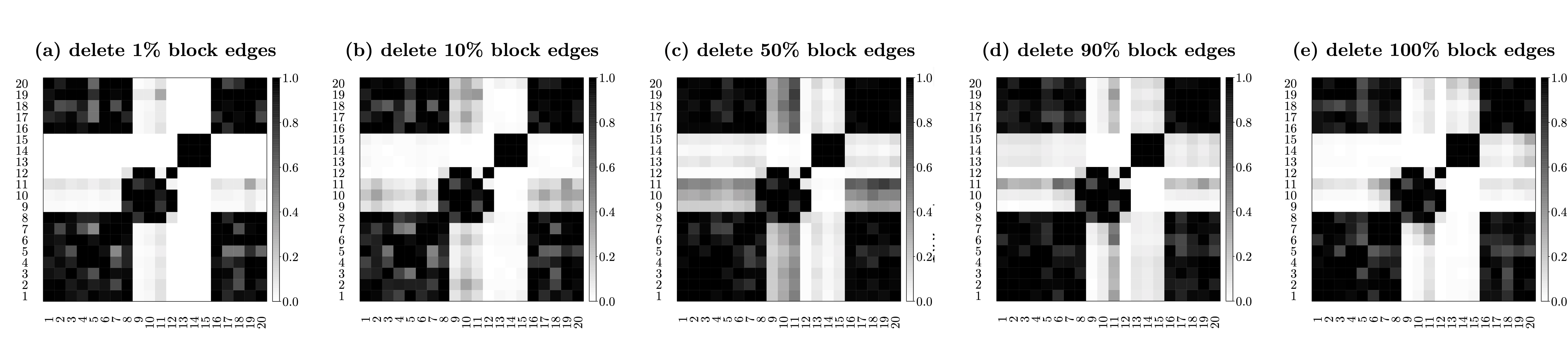}}
\caption*{\it Figure S6: Sensitivity analysis Case 2: delete blocks of edges}
\label{fig:supplSensitivityII}
\end{figure}


\section*{S7: Prior specifications in the GDSC data analysis}
\subsection*{}

To specify weak informative priors for the random effects for the 13 cancer tissue types, we set the hyper-parameter $w_0 \sim  \mathcal{IG}(54.6, 400)$ by assuming each cancer tissue effect with prior mean 0 and prior standard deviation 3 and according to the prior specification in section S5. Corresponding to the three feature sets, the model sparsity is assumed to be 6\%, 2\% and 1\% respectively
\footnotemark
\footnotetext{We assume there are roughly 50 associated features for each response variables based on the number of drug target genes. Then the assumed model sparsity is 50 divided by the total number of features, i.e., 50/763, 50/1699 and 50/3096.}
, so we obtain parameter $d$ of the MRF prior with values -2.7, -4 and -4.6. Our SSUR-MRF model with random effects, hereafter denoted as SSUR-MRF, has good MCMC diagnostics (see Section S8). For comparison, we run a similar model with parameter $e=0$ of the MRF prior, which is equivalent to independent Bernoulli priors for the latent indicator variables, here denoted as SSUR-Ber. In SSUR-Ber, the model sparsity hyper-parameter $d$ in the prior $f(\bm{\gamma}|d) \propto \exp\{d\mathbbm{1}^\top\bm{\gamma}\}$ is tuned to reach the best elpd$_{\text{loo}}$ and elpd$_{\text{waic}}$.


\clearpage
\section*{S8: MCMC diagnostics by SSUR-Ber and SSUR-MRF in the GDSC data analysis}
\subsection*{}
\vspace{-6mm}
\begin{figure}[H]
\centering
\includegraphics[height=0.56 \textwidth]{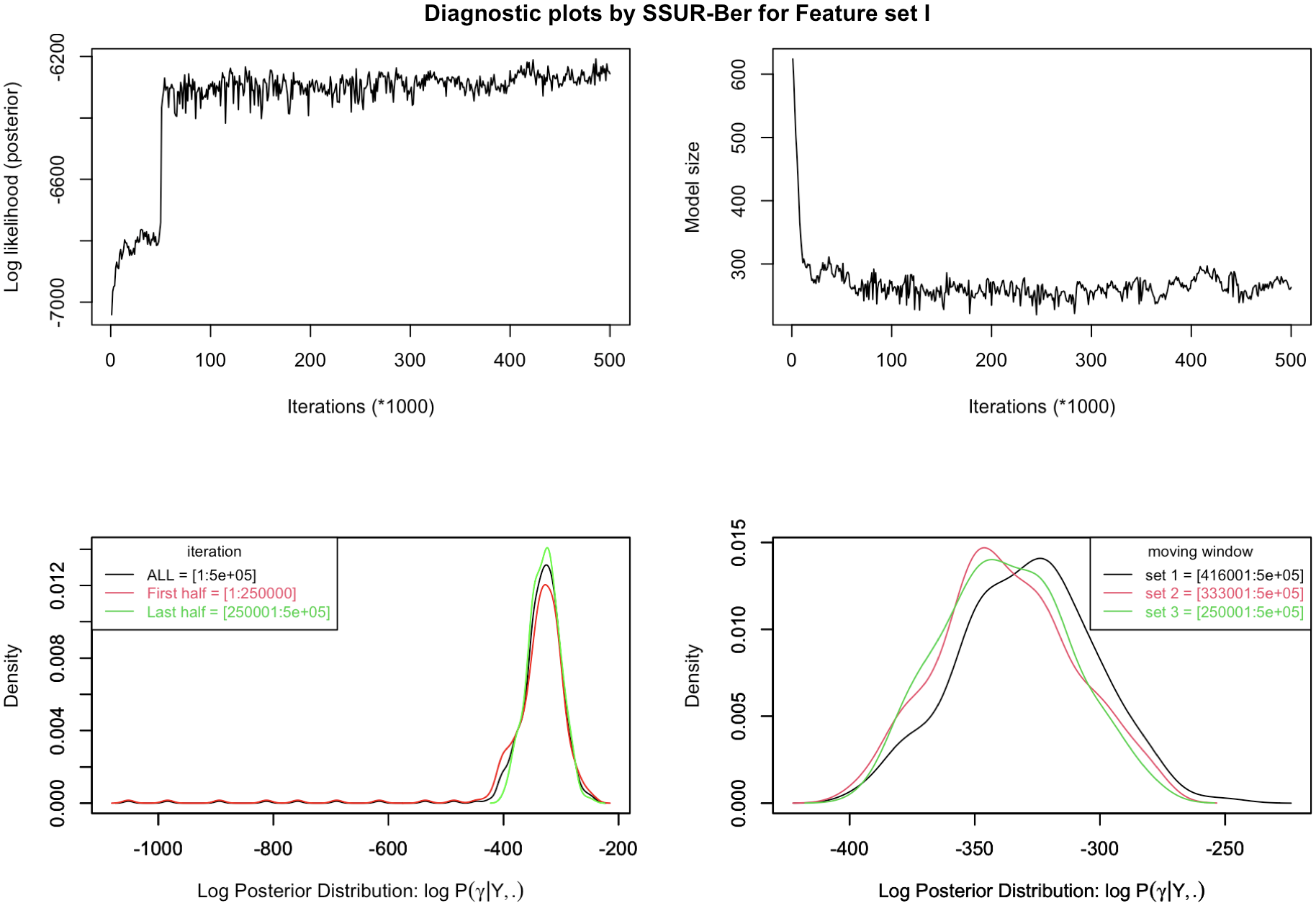}
\caption*{\it Figure S8.1.1: Diagnostic plots. The hyperparameters set \texttt{hyperpar = list({\bf mrf\_d=-2, mrf\_e=0}, a\_w0=55, b\_w0=400, a\_w=4, b\_w=32)}, and others are by default.}
\end{figure}

\vspace{-4mm}
\begin{figure}[H]
\centering
\includegraphics[height=0.56 \textwidth]{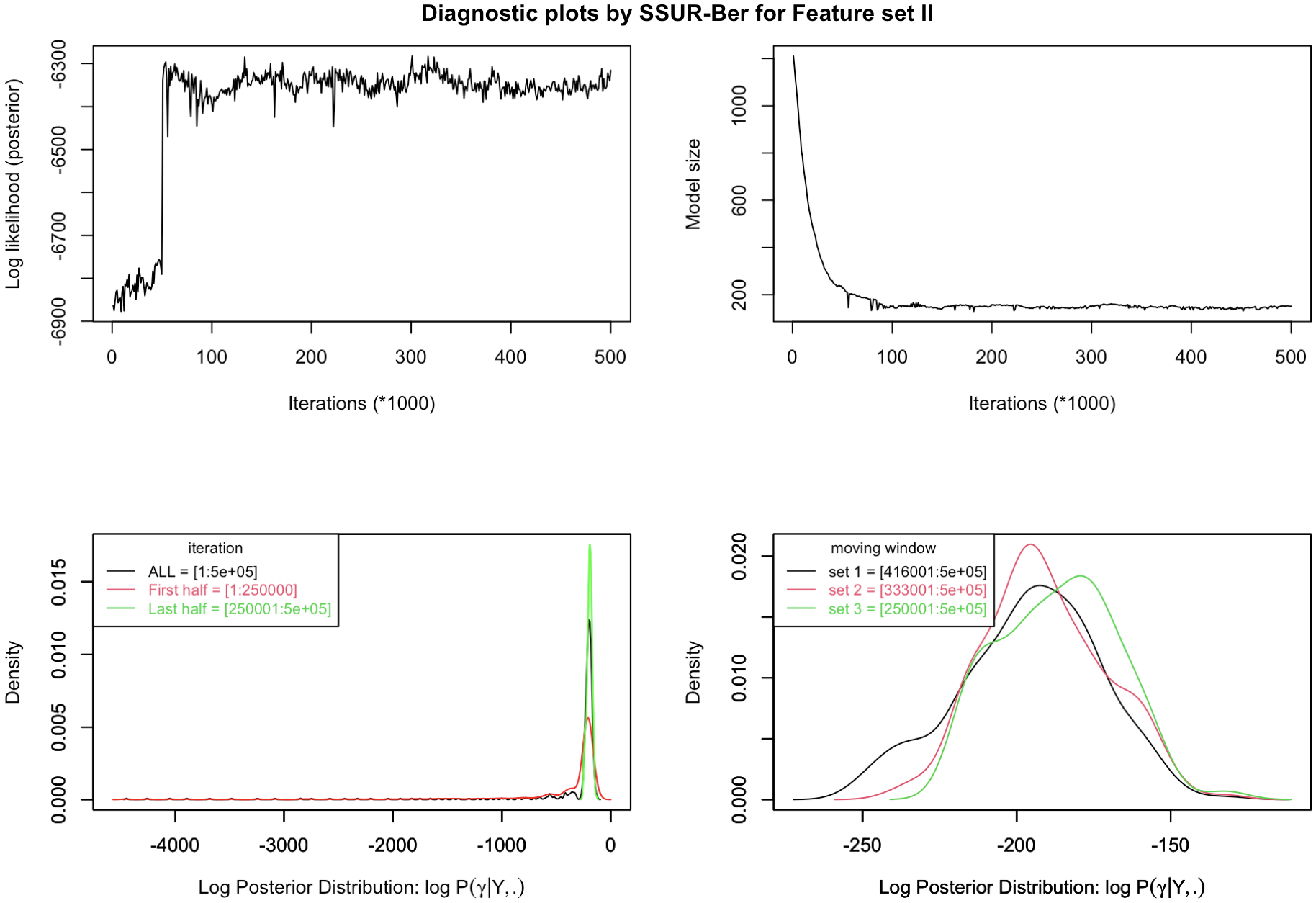}
\caption*{\it Figure S8.1.2: Diagnostic plots. The hyperparameters set \texttt{hyperpar = list({\bf mrf\_d=-4, mrf\_e=0}, a\_w0=55, b\_w0=400, a\_w=4, b\_w=32)}, and others are by default.}
\end{figure}

\begin{figure}[H]
\centering
\includegraphics[height=0.56 \textwidth]{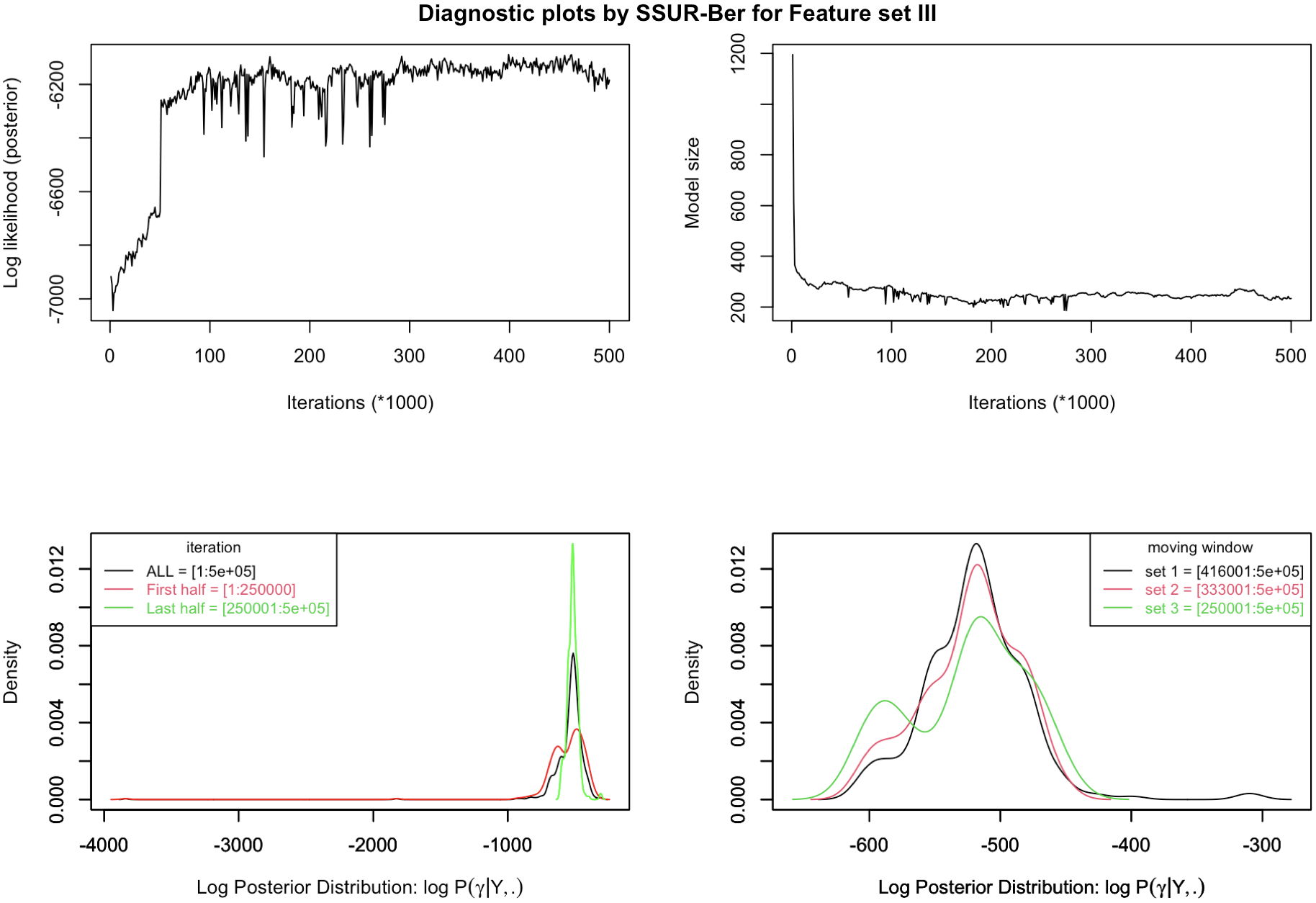}
\caption*{\it Figure S8.1.3: Diagnostic plots. The hyperparameters set \texttt{hyperpar = list({\bf mrf\_d=-3.5, mrf\_e=0}, a\_w0=55, b\_w0=400, a\_w=4, b\_w=33)}, and others are by default.}
\end{figure}

\begin{figure}[H]
\centering
\includegraphics[height=0.56 \textwidth]{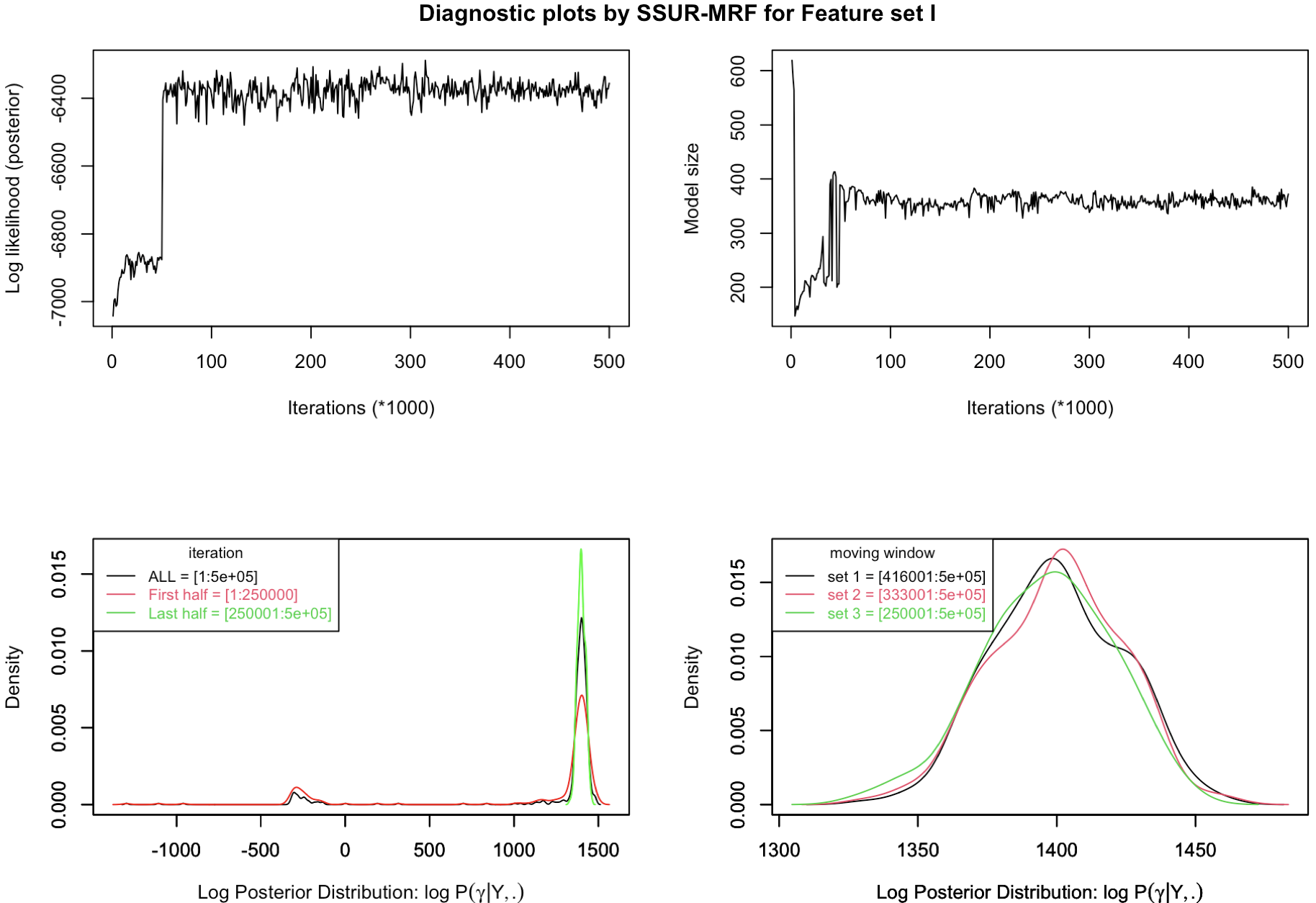}
\caption*{\it Figure S8.2.1: Diagnostic plots. The hyperparameters set \texttt{hyperpar = list({\bf mrf\_d=-2, mrf\_e=0},  a\_w0=55, b\_w0=400, a\_w=4, b\_w=32)}, and others are by default.}
\end{figure}

\begin{figure}[H]
\centering
\includegraphics[height=0.56 \textwidth]{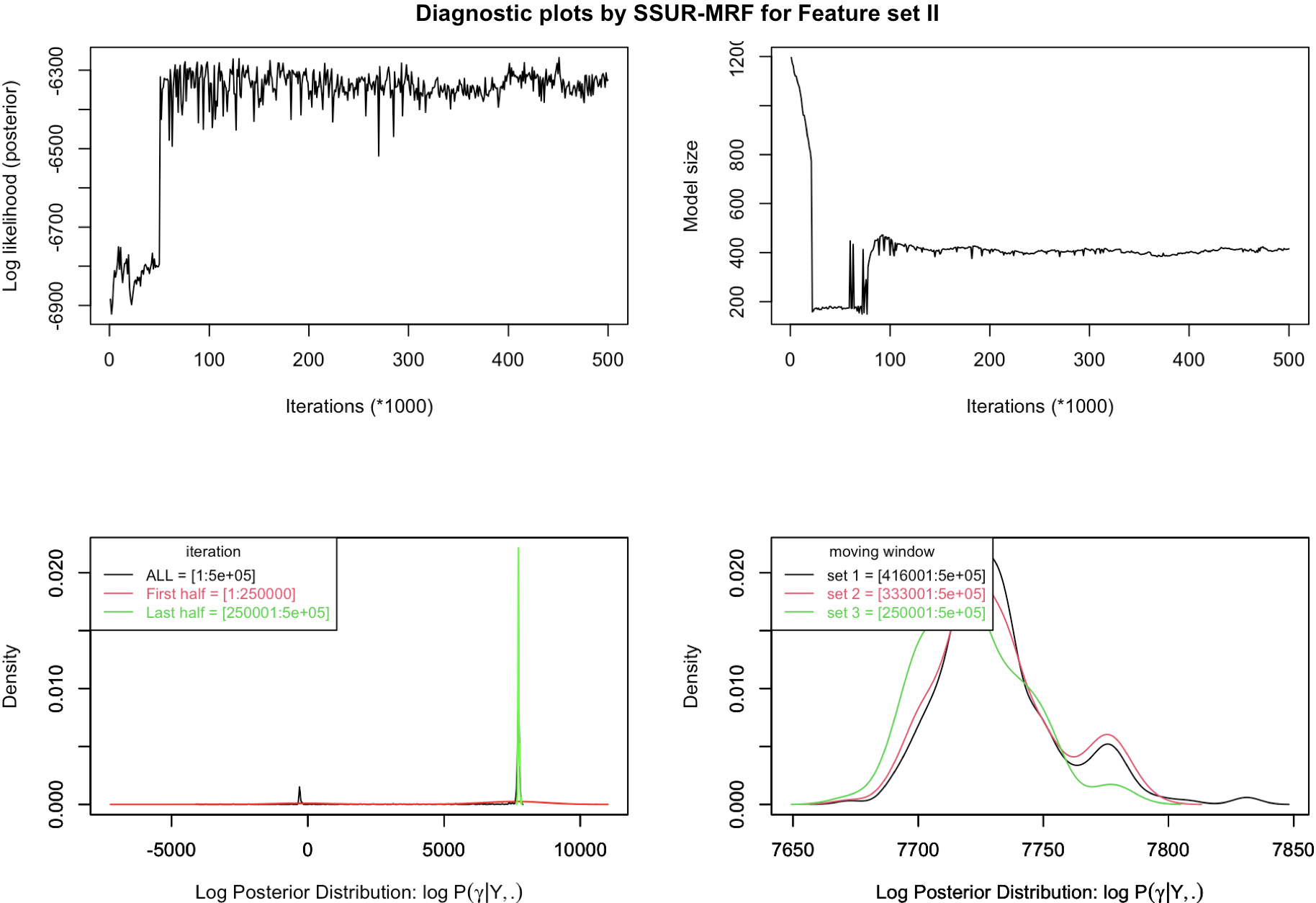}
\caption*{\it Figure S8.2.2: Diagnostic plots. The hyperparameters set \texttt{hyperpar = list(mrf\_d=-4, mrf\_e=0.3, a\_w0=55, b\_w0=400, a\_w=4, b\_w=32)}, and others are by default.}
\end{figure}

\begin{figure}[H]
\centering
\includegraphics[height=0.56 \textwidth]{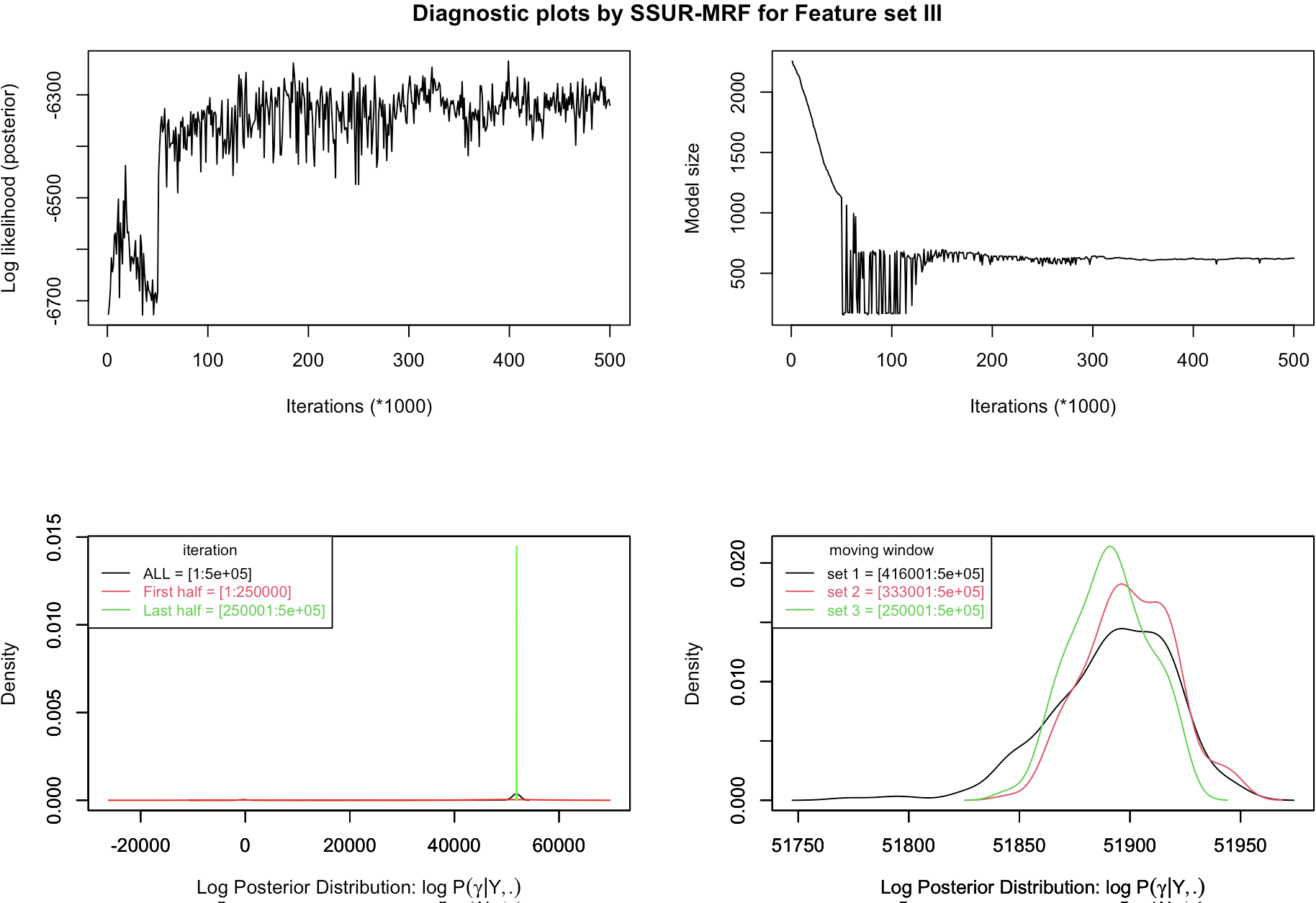}
\caption*{\it Figure S8.2.3: Diagnostic plots. The hyperparameters set \texttt{hyperpar = list(mrf\_d=-4.6, mrf\_e=0.5, a\_w0=55, b\_w0=400, a\_w=4, b\_w=33)}, and others are by default.}
\end{figure}
\vspace{-2mm}

\clearpage
\section*{S9: Estimated residual structure between the seven drugs in the GDSC data analysis}
\subsection*{}

\begin{figure}[H]
\centering
\includegraphics[height=0.8 \textwidth]{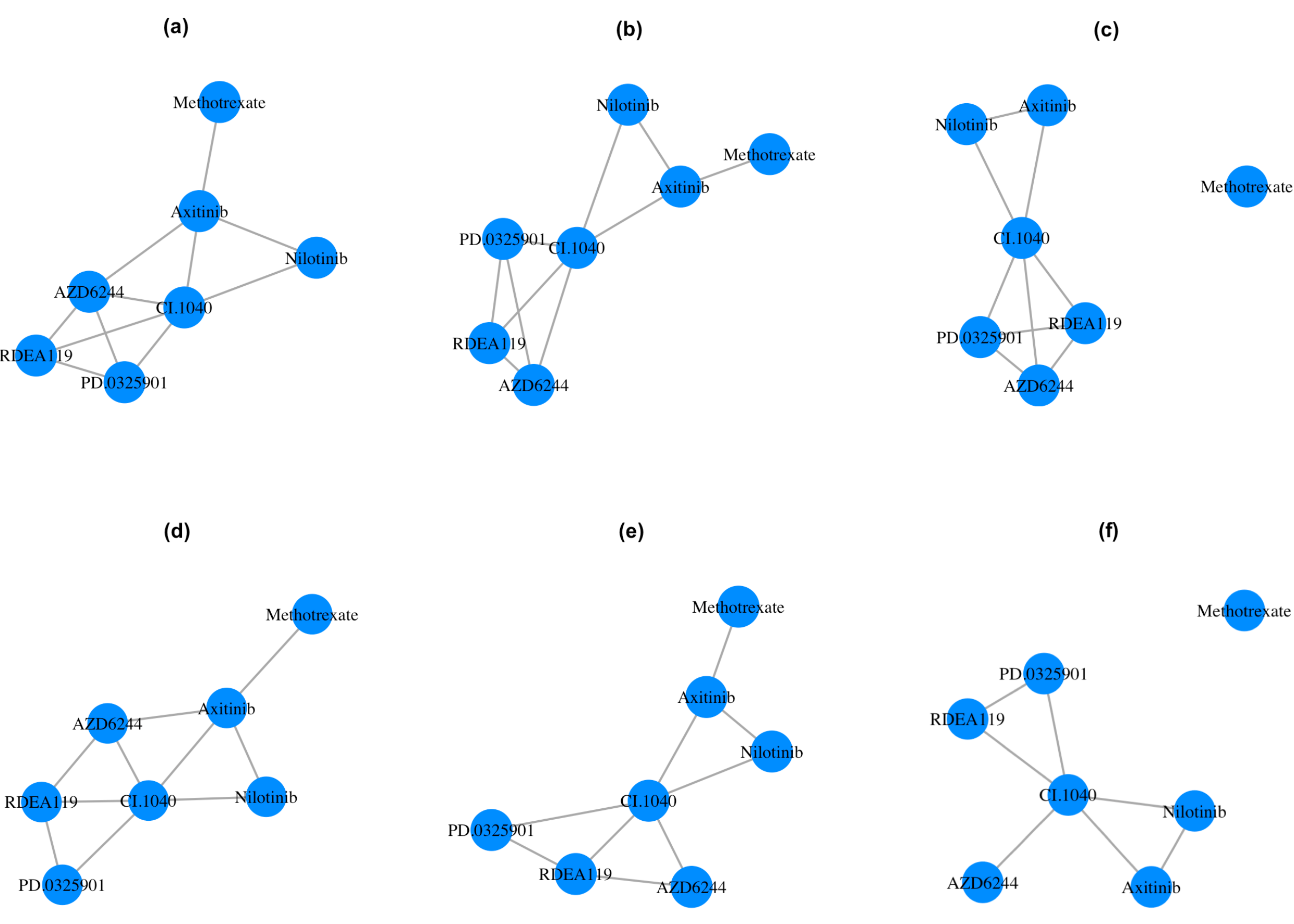}
\caption*{\it Figure S9: GDSC data application: Estimated residual structure between the seven drugs by SSUR-MRF and SSUR-Ber based on $\hat{\mathcal{G}}$ thresholded at 0.5. Panels (a)-(c) are estimated by SSUR-MRF, and panels (d)-(f) are estimated by SSUR-Ber,  corresponding to Feature sets I, II and III respectively}
\label{figure:Gy-MRF}
\end{figure}

\clearpage
\section*{S10: Identified genomic features for the MAPK inhibitors by SSUR-Ber in the GDSC data analysis}
\subsection*{}

\begin{table}[H]
\centering
\caption*{\it Table S10.1: GDSC data application: Identified genomic features for the MAPK inhibitors by SSUR-Ber } 
\label{tab:gdsc-bernoulli}
{
\begin{tabular}{ll}
\hline\hline
& \multicolumn{1}{c}{Identified genomic features}\\
\hline
Feature set I & \texttt{ABCC3; LXN; LGALS5.GALIG; MAGED4.LOC653210; ZNF198.CNV} 
\smallskip\\
Feature set II & \texttt{IGFBP6; CA14; C1ORF115; LAMC2; TIMP1 } 
\smallskip\\
Feature set III & \texttt{ATF5; FAAH; ABCB4; IGFBP6; IGFBP2; CENTD3; MAST4; MITF;}\\ 
& \texttt{FGR; CA14; PYGL; MEIS1; LAMC2; SPINK1; CRISPLD2; FLT3;}\\ 
& \texttt{P2RY10; ZNF204; GRAMD3; ELMO1; LIFR; ALDH3B1; SPRY2;}\\ 
& \texttt{S100A11; KCTD12; PBX1.CNV; FBXW7.MUT; BCR\_ABL.MUT} \\
\hline\hline
\end{tabular}
}
\end{table}

\begin{table}[H]
\centering
\caption*{\it Table S10.2: GDSC data application: Identified common features across feature sets I, II, and III for the MAPK inhibitors by SSUR-MRF.} 
\label{tab:gdsc-common-genes}
{
\begin{tabular}{ c }
\hline\hline
Identified common features\\
\hline
\texttt{AKT2.CNV; ALK.CNV; BRCA1.CNV; DDX10.CNV; EIF4A2.CNV; ELL.CNV;}\\    
\texttt{FGFR1OP.CNV; FGFR3.CNV; FH.CNV; IDH1.CNV; KTN1.CNV; MDM2.CNV;}\\   \texttt{MYCL1.CNV; NF2.CNV; NIN.CNV; NSD1.CNV; NTRK3.CNV; PDGFRA.CNV;}\\   \texttt{PDGFRB.CNV; PER1.CNV; RANBP17.CNV; TPM3.CNV; ALK.MUT; BRCA1.MUT;}\\     
\texttt{EP300.MUT; FGFR3.MUT; FLCN.MUT; IDH1.MUT; MAP2K4.MUT; MDM2.MUT;}   \\
\texttt{MYCL1.MUT; NF2.MUT; PDGFRA.MUT; PIK3CA.MUT; TSC1.MUT}   \\
\hline\hline
\multicolumn{1}{p{15cm}}{NOTE: Gene RANBP17.CNV is not in the Cancer Gene Census database.} \\
\end{tabular}
}
\end{table}
  
\begin{table}[H]
\centering
\caption*{\it Table S10.3: GDSC data application: Identified features by Feature set III but not identified by Feature set I or II for the MAPK inhibitors by SSUR-MRF.} 
\label{tab:gdsc-common-genes}
{
\begin{tabular}{ c }
\hline\hline
Identified features by Feature set III but not identified by Feature set I or II\\
\hline
\texttt{PLEKHB1; PTPRU; RGC32; ARNT2; SLC30A3; PTPN6; SNTG2; IL1R1; PLTP;} \\
\texttt{TRPS1; CCND2; EFS; PTK2B; MMP9; GNG7; CA14; BCAR3; ZNF532; HSPA6;} \\ 
\texttt{TMEM45A; DUSP4; DUSP2; DUSP1; HPGD; IL18; IL24; NRTN; FGFR3; TRDV2;} \\
\texttt{USH1C; CDO1; TIAM1; ATP2A3; NLGN1; LOC284244; FLJ14082; LGALS3;} \\    
\texttt{CYP26A1; BCL2; RASGRP1; RASGRP2; SYNPO; RORA; TRAC; KRT6B; ACTN1;} \\ 
\texttt{ZFP30; PRKCB1; PDGFRB; HOXA10; KHDRBS3; RIBC2; LCN2; TSPY1;}\\
\texttt{CDC42EP1; MYH10; WIF1; CCND2.CNV; KRAS.CNV; CCND2.MUT; KDM5C.MUT}\\
\hline\hline
\end{tabular}
}
\end{table}

\clearpage
\section*{S11: Prediction performance for Feature sets I, II and III in the GDSC data analysis}
\subsection*{}

\begin{table}[H]
\centering
\caption*{\it  Table S11: GDSC data application: Similar prediction performance between the SSUR-Ber and SSUR-MRF models.} 
\begin{tabular}{rcc}
\hline
\hline
\smallskip
 & SSUR-Ber & SSUR-MRF \\ \cline{1-3}
{\bf Feature set I} \\
elpd.LOO & -8190 & -8162 \\
elpd.WAIC & -8219 & -8212\\
RMSE & 2.116 & 2.133 \\
RMSPE & 2.110 & 2.097 \\
\smallskip
{\bf Feature set II} \\
elpd.LOO & -8135 & -8140 \\
elpd.WAIC & -8136 & -8139 \\
RMSE & 1.866 & 1.810 \\
RMSPE & 1.810 & 1.972 \\
\smallskip
{\bf Feature set III} \\
elpd.LOO & -8136 & -8143 \\
elpd.WAIC & -8169 & -8178 \\
RMSE & 2.003 & 1.883 \\
RMSPE & 2.095 & 2.062 \\
\hline
\hline
\end{tabular}
\end{table}

\clearpage
\section*{S12: Drug responses across cancer tissue types for each drug}
\subsection*{}

\begin{figure}[H]
\centering
\includegraphics[height=1.1 \textwidth]{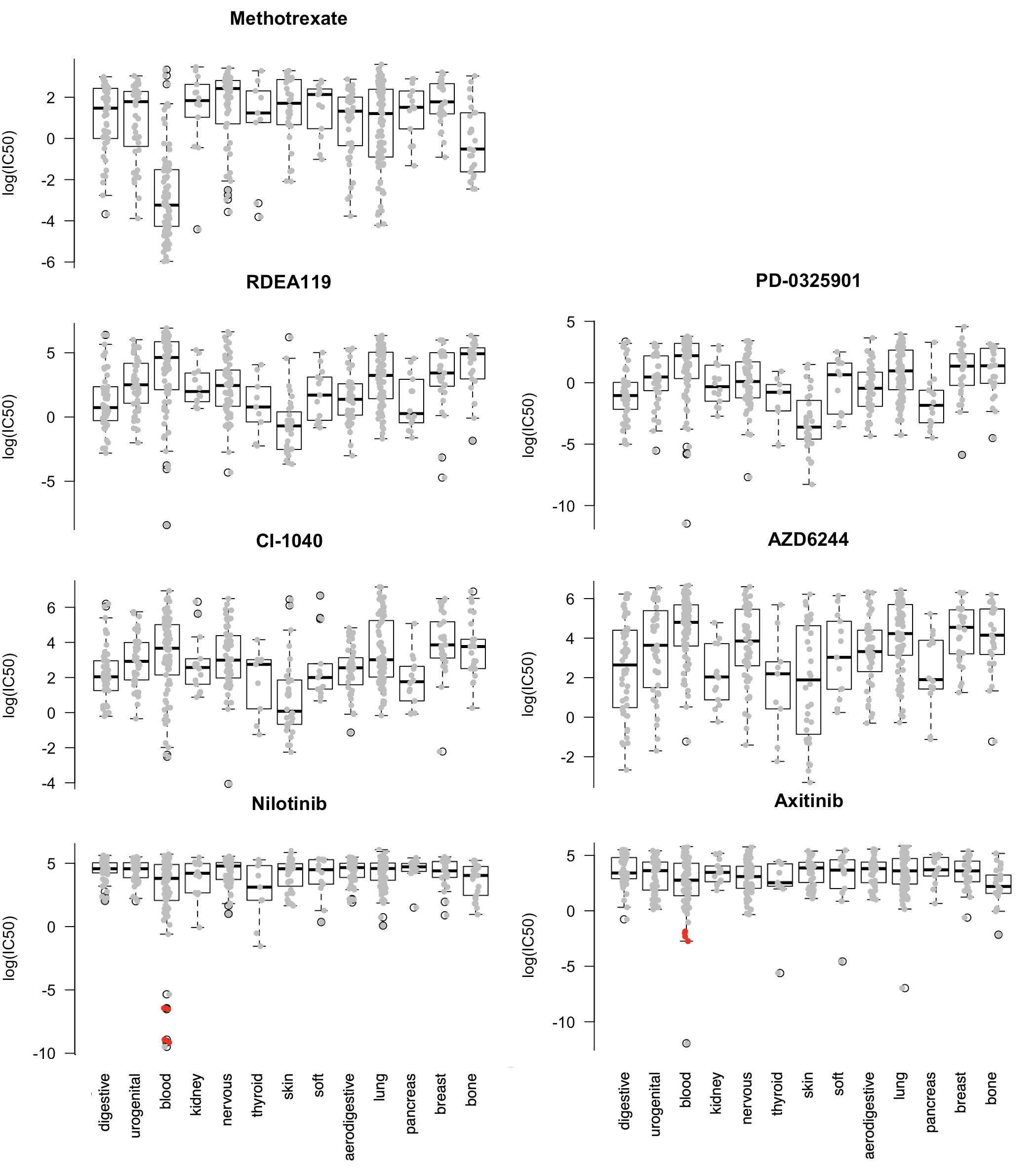}
\caption*{\it Figure S12: GDSC data application: Drug responses across cancer tissue types for each drug. Each grey dot corresponds each cancer cell line. Red points ``$\textcolor{red}{\bullet}$" correspond to the BCR-ABL mutated blood cell lines treated by drugs Nilotinib and Axitinib.}
\end{figure}

\end{document}